\documentclass[reprint,aps,a4paper,prd,floatfix,twocolumn,amsmath,amssymb,nofootinbib]{revtex4-1}
\usepackage{lipsum}
\usepackage[utf8]{inputenc}
\usepackage{amsmath,ulem}
\usepackage{cases,color}
\usepackage{indentfirst}
\usepackage{natbib}
\usepackage{appendix}
\usepackage[T1]{fontenc}
\usepackage{bbold}
\usepackage[table]{xcolor}

\usepackage{mathrsfs}
\usepackage{amssymb}
\usepackage{multirow}

\usepackage{slashed}

\usepackage{braket}

\usepackage{graphicx}
\usepackage{float}

\usepackage{cases}

\begin{document}
\title{
Two-solar-mass hybrid stars: A two model description using the Nambu-Jona-Lasinio quark model}

\author{Renan Câmara Pereira}
\email{renan.pereira@student.fisica.uc.pt}
\affiliation{CFisUC, Department of Physics,
University of Coimbra, P-3004 - 516  Coimbra, Portugal}

\author{Pedro Costa}
\email{pcosta@teor.fis.uc.pt}
\affiliation{CFisUC, Department of Physics,
University of Coimbra, P-3004 - 516  Coimbra, Portugal}

\author{Constança Providência}
\email{cp@teor.fis.uc.pt}
\affiliation{CFisUC, Department of Physics,
University of Coimbra, P-3004 - 516  Coimbra, Portugal}

\date{\today}

\begin{abstract}
Hybrid stars with a quark phase described by the Nambu$-$Jona-Lasinio  model are studied. The hadron-quark model used to determine the stellar matter equation of state favors the appearance of quark matter: the coincidence of the deconfinement and  chiral transitions and a low vacuum constituent quark mass. These two properties are essential to build equations of state that predict pure quark matter in the center of neutron stars.  The effect of vector-isoscalar and vector-isovector terms is discussed, and it is shown that the  vector-isoscalar terms are necessary to describe 2$M_\odot$ hybrid stars, 
and the vector-isovector terms result in larger quark cores and a smaller deconfinement density.
\end{abstract}

\maketitle


\section{Introduction}

Compact stars are natural laboratories to investigate the properties of strongly interacting matter at high densities and small temperatures. 
Due to their very large central densities, several times larger than normal saturation density, it is possible that the deconfinement phase transition and the partial restoration of chiral symmetry may occur inside compact stars. 
Indeed, as density increases baryons start to overlap, the distance between quarks becomes very short, and distinct baryons gradually cease to exist. Consequently, inside a compact star the density could be high enough to involve quark degrees of freedom. The study of the behavior of the matter under extreme conditions such as the ones existing in the interior of neutron stars, should take into account that at low densities the relevant degrees of freedom are hadrons while at high densities quark degrees of freedom may set in giving rise to hybrid stars. 
However, the two solar mass pulsars PSR J0348+0432 ($M=2.01\pm$0.04 $M_\odot$) \cite{Antoniadis:2013pzd} and  PSR J1614-2230 (with the recently updated mass 1.928$\pm 0.017$ $M_\odot$ \cite{Demorest:2010bx,Fonseca:2016tux}) set a strong constraint on the high density equation of state (EoS), in particular, 
on the possible existence of hyperons, kaon condensation or even quark matter inside neutron stars.

In Ref. \cite{Glendenning} hybrid stars are described  using a two model approach,  hadronic matter within a nuclear field theory model and quark matter within  the MIT bag model \cite{Chodos:1974pn}. The hadron-quark phase transition is obtained imposing Gibbs conditions and considering  global electric charge neutrality, which develops a mixed phase separating a pure hadronic and a pure quark phase. The MIT bag model is a quite simple model that has been widely used. A quark core is possible if the bag constant is not too high. This parameter is constrained from below, imposing that at saturation density nuclear matter  has a lower energy than strange matter, see Ref. \cite{Bombaci:2004mt}.  

In Ref. \cite{Schertler:1999xn}, the authors consider, instead of the MIT bag model, the SU(3) Nambu$-$Jona-Lasinio (NJL) model to describe the quark phase.  It is shown that a pure quark phase does not occur inside a neutron star, although quarks might exist as part of a non-homogenous quark-hadronic mixed phase in the center of the star, in stars with a mass close to the maximum allowed mass, $\sim1.7\, M_\odot$. As in \cite{Glendenning}, hadronic matter is described within a relativistic mean field (RMF) model. Similar results are obtained applying a Brueckner Hartree-Fock approach to describe the hadronic phase, and even if a superconducting quark phase is considered for the quark phase  within the NJL model \cite{Baldo:2002ju}. Maximum mass stars obtained  have a mass $\sim$ 1.8 $M_\odot$, becoming unstable as soon as quark matter sets in. At  finite temperature \cite{Menezes:2003xa} it was possible to obtain pure quark matter in the star center describing quark matter within NJL model but masses below 1.9 $M_\odot$ were obtained.

However, contrary to \cite{Baldo:2002ju},  a stable cold hybrid star with a diquark condensation in the quark phase was obtained in \cite{Shovkovy:2003ce} within a SU(2) NJL model. This different behavior was attributed  in \cite{Buballa:2003et},  to the different vacuum constituent quark masses obtained in both calculations, and, in particular,   it was shown that the hadron-quark phase transition is controlled by the constituent mass of the nonstrange quarks in vacuum, and that  smaller vacuum constituent masses favor the appearance of a pure quark phase because the zero pressure is shifted to smaller chemical potentials. 

A stable cold quark phase  has also been obtained within SU(3) NJL model by the introduction of a bag constant, $B^*$, which guarantees that the partial restoration of chiral symmetry coincides with the transition from hadronic to the quark matter \cite{Pagliara:2007ph}. This constant shifts the effective bag constant as defined in \cite{Buballa:1998pr} to smaller values and favors the hadron-quark transition.
However,  no two solar mass hybrid stars were predicted. In Ref. \cite{Bonanno:2011ch} the fixing condition of the bag constant $B^*$ was relaxed and the deconfinement baryonic density, which was chosen beforehand, was used to determine the bag constant. Stars with over two solar masses and a quark core in a color super-conducting phase were obtained with a vector interaction added to the NJL Lagrangian density.  In Ref. \cite{Logoteta:2013ipa} the consequences of quark nucleation were studied and it was shown that not all two solar mass hybrid star configurations are populated after nucleation. 

Multiple other studies of the quark-hadron phase transition in neutron stars, involving several approaches for the description of the quark matter and the hadronic matter have been performed. The topical issue \cite{EPJA2016} includes several articles that review different aspects of this problem. Some other approaches used to describe quark matter are  the  field correlator method \cite{Burgio:2015zka},  perturbative QCD \cite{Fraga:2015xha}, the chromo-dielectric model \cite{Drago:1995pr,Logoteta:2012zz}, also a new class of two-phase EoS for hybrid stars was discussed in \cite{Alvarez-Castillo:2016oln}. In \cite{Lawley:2006ps}, a unified approach to the EoS of a hybrid star was proposed with
both nuclear and quark matter described within the framework of the NJL model, and, moreover, the internal quark structure of the free nucleon was taken into account. However, stable hybrid stars were only possible with a quite strong pairing interaction and maximum masses below 1.5 $M_\odot$ were obtained.  More recently in \cite{Pais:2016dng}, hybrid stars were also described 
in the framework of the NJL model for both the hadronic and the quark phases, but structureless nucleons were considered  in the hadronic phase and the couplings were fitted independently  in each one of the phases, contrary to \cite{Lawley:2006ps}. However,  in \cite{Pais:2016dng}, the nucleonic EoS satisfies experimental and theoretical constraints at subsaturation, saturation and suprasaturation densities and 2$M_\odot$ stable hybrid stars have been obtained.

The role of the vector interaction, responsible for the excitations of vector and pseudovector mesons, in the properties of compact stars has been extensively studied within the SU(3) NJL model (see for example \cite{Hanauske:2001nc,Klahn:2006iw,Pagliara:2007ph,Bonanno:2011ch,Lenzi:2012xz,Masuda:2012ed,Klahn:2013kga,Logoteta:2013ipa,Pais:2016dng,Menezes:2014aka,Klahn:2015mfa,Ferrer:2015vca}).
It is known that for a positive $G_V$ the vector interaction provides a repulsive interaction between quarks. This aspect is very important because it stiffens the NJL EoS, which is essential to describe high-mass hybrid stars. Models with a larger $G_V$ give larger maximum star masses \cite{Bonanno:2011ch,Lenzi:2012xz}.   

Concerning the effect of the vector interaction on the QCD phase diagram, namely on the chiral first-order transition, it has been shown that when $G_V$ is positive (negative) it contributes to weaken (strengthen) the first-order transition due to repulsive (attractive) nature of the interaction \cite{Fukushima:2008wg}. 
Indeed, a repulsive interaction shrinks the first-order transition region, which forces the critical end point to occur at smaller temperatures, and as $G_V$ increases the first-order transition occurs at higher baryonic chemical potentials.

However, in spite of its importance, the value of the vector coupling, $G_V$, has not yet been definitively settled: its value in the vacuum can be determined by fitting the vector meson spectrum \cite{Klimt:1989pm,Lutz:1992dv} but it is not evident that the value of $G_V$ in the medium has to be the same as in the vacuum \cite{Fukushima:2008wg}. 
In fact, finite-density environment might give rise to a vector interaction, described by a finite $G_V$, even though the contribution of this interaction is zero in the vacuum \cite{Fukushima:2008wg}.
On the other hand, recent studies of the QCD phase diagram using the { extended version of the NJL model with Polyakov loop} suggest that the magnitude of $G_V$ may be comparable to or larger than the coupling $G_S$ \cite{Bratovic:2012qs,Lourenco:2012yv}, so as  most works we will also consider $G_V$ as a free parameter and vary its magnitude in the range $0\leq G_V/G_S \leq 1$.

The main objective of the present work is to study the possibility of obtaining two solar mass hybrid stars with a quark core described within the NJL model considering a more generalized interaction than the one used in  previous works.  The hadronic sector will be described within an RMF model.   We will perform a complete study considering both the  SU(2) and SU(3) NJL versions, however, only  the latter allows the inclusion of strangeness which will probably exist inside compact stars.
In fact, it is expected that in the interior of a neutron star strangeness will be present either in the form of hyperons,  kaon condensation or deconfined quark matter \cite{Glendenning}.

Previous studies have shown that a quark phase is favored if a smaller vacuum constituent quark mass than the one obtained with the SU(3) NJL parametrization given in \cite{Rehberg:1995kh} is used and if  a bag constant $B^*$ is included.  We will, therefore, investigate how the choice of the hadron and quark EoS, obtained from two independent models, one for the hadronic phase and another for the quark phase, for the calculation of a hybrid star EoS, allows the description of 2$M_\odot$ stars. In particular, we will consider: 
a) a low vacuum constitutent quark mass; 
b) that the deconfinement phase transition coincides with the partial restoration of chiral symmetry. The first condition is implemented by fitting the NJL parameters including a constraint on the vacuum constituent mass and the second by introducing  an effective bag constant, $B^*$, which guarantees that the chiral symmetry transition\footnote{In the present work chiral symmetry transition refers to the transition to the phase where chiral symmetry is partially restored.} coincides with the transition from the hadronic to the quark matter \cite{Pagliara:2007ph}. We will analyze the effect of the vector interaction in the properties of compact stars under the conditions described above.

This work is structured as follows. 
In Sec. \ref{formalism} we present the EoS for hadronic matter used at low densities, and the EoS for quark matter obtained within the SU(3) NJL model including vector interaction [the SU(2) model is also presented for comparison purposes]. We also discuss the conditions of matter in $\beta$-equilibrium and the Gibbs phase equilibrium conditions together with the procedure used to fix the effective bag constant, $B^*$, within NJL models.
Section \ref{results} is devoted to present the results for the possible existence of hybrid stars within the SU(2) NJL model and within the extension to the SU(3) NJL model in order to take into account strangeness.
Finally, Sec. \ref{conclusions} is dedicated to concluding remarks.


\section{Formalism}
\label{formalism}

In order to perform our investigation, quark matter is described by the NJL model, in both SU(2) and SU(3) versions, with vector interactions. The SU(3) NJL model will  allow us to explore the influence of strangeness in the quark EoS. 
Indeed, for densities above $\sim 2-3 \rho_0$ there is enough energy in the system for strangeness to become relevant. The central densities inside a neutron star are well above this value, therefore, a model that includes strangeness represents a more realistic study of these systems.
Comparing both SU(2) and SU(3) versions of the NJL model will allow us to infer the role of strangeness in the system.

Hadron matter is described by a RMF nuclear model. To describe the mixed phase we impose local electric charge neutrality and the Gibbs criteria: the pure hadronic phase and the quark phase are connected to each other through  mechanical, thermal and chemical equilibrium.


\subsection{Hadronic matter}
\label{hadronicEOS}

The relativistic mean-field model NL3$\omega\rho$ \cite{Horowitz:2000xj,Fortin:2016hny} will be used to  describe the hadronic (confined) phase of the system in $\beta-$equilibrium. The Lagrangian density of the model reads
\begin{widetext}
\begin{align}
\mathscr{L}&=\sum_{N=p,n} 
\bar{\psi}_N\left[\gamma^{\mu}(i\partial_{\mu}-g_{\omega N}\omega_{\mu
}-\frac{1}{2} g_{\rho N} {\boldsymbol \tau} \cdot {\boldsymbol 
\rho_{\mu}} )\nonumber -(m_{N}-g_{\sigma N}\sigma)\right]\psi_N \nonumber\\
&+\frac{1}{2}\partial^{\mu}\sigma\partial_{\mu
}\sigma-\frac{1}{2}m_{\sigma}^{2}\sigma^{2}  
- \frac{1}{4} \Omega_{\mu \nu} \Omega^{\mu \nu}+ 
\frac{1}{2}m_{\omega}^{2}\omega^{\mu}\omega_{\mu}-\frac{1}{4}{\boldsymbol
   \rho}^{\mu\nu} \cdot {\boldsymbol 
\rho}_{\mu\nu}+\frac{1}{2}m_{\rho}^{2}{\boldsymbol
   \rho}^{\mu}\cdot{\boldsymbol \rho}_{\mu}\nonumber \\
&-\frac{1}{3}b m_{N} (g_{\sigma N} \sigma)^3 
-\frac{1}{4}c (g_{\sigma N} \sigma)^4 
+\Lambda_{\omega}\left(g^{2}_{\omega} 
\omega_{\mu}\omega^{\mu}\right)\left(g^{2}_{\rho} \boldsymbol{\rho}_{\mu}
\cdot\boldsymbol{\rho}^{\mu}\right).
\label{Lag_hadron}
\end{align}
\end{widetext}
This model contains several nonlinear terms:  besides the usual cubic and quartic terms on the $\sigma$-meson, there is also a  quartic term that mixes the $\omega$ and the $\rho$-meson and which results in a softening of the symmetry energy at large densities. 
However, since it does not include a quartic term on the $\omega$-meson it has a quite stiff EoS at large densities. No hyperons are included in the present study. The onset of hyperons will certainly compete with the quark onset.
But, as shown in Ref. \cite{Fortin:2016hny} the onset of hyperons for NL3$\omega\rho$ occurs at 0.31 fm$^{-3}$, above the onset of quark matter as we will see in Sec. \ref{res_SU(3)}. Therefore,  we will only consider nucleonic matter in the hadronic phase because the appearance of hyperons  in some cases only would make the comparisons difficult.

The NL3$\omega\rho$ model has the following saturation properties (see 
\cite{Horowitz:2000xj,Fortin:2016hny}): saturation density 
$\rho_0=0.148$ fm$^{-3}$, binding energy $E/A= -16.30$ MeV, 
incompressibility $K= 271.76$ MeV, symmetry energy $J=31.7$ MeV, 
symmetry energy slope $L=55.5$ MeV and effective mass $M^*/M= 0.60$.  In 
\cite{Fortin:2016hny} it was shown that this model satisfies a 
reasonable amount of constraints: experimental, astrophysical and theoretical from 
microscopic neutron matter calculations. In particular, the maximum 
possible neutron star mass is 2.75 $M_\odot$, well above the 2$M_\odot$ 
constraint imposed by the pulsars J1614-2230 and J0348+043.

\subsection{The NJL model}
\label{SU3}

The quark phase of the EoS is described within the SU(3) NJL model including, besides the four quark interaction and the 't Hooft determinant that breaks the $U_A(1)$ symmetry, vector and pseudovector terms (both vector-isoscalar and vector-isovector will be considered).

The Lagrangian density is written as,
\begin{widetext}
\begin{align}
\mathscr{L}&=\bar{\psi}(i\slashed{\partial}+\hat{m}+\gamma^0\hat{\mu})\psi  
+ G_S  \sum\limits_{a=0}^8 \left[ \left(\bar{\psi}\lambda^a\psi \right)^2 + \left(\bar{\psi}i\gamma_5\lambda^a\psi \right)^2  \right]\nonumber\\
&- G_D\left[ \det \left(\bar{\psi}(1+\gamma_5)\psi \right) + \det \left(\bar{\psi}(1-\gamma_5)\psi \right)  \right] - \mathscr{L}_{vec},
\label{Lag_quark}
\end{align}
with,
\begin{align}
\mathscr{L}_{vec}&=
     G_\omega\Big[ (\bar{\psi}\gamma^\mu\lambda^0\psi)^2 + (\bar{\psi}\gamma^\mu\gamma_5\lambda^0\psi)^2 \Big] +
     G_\rho\sum\limits_{a=1}^8\Big[ (\bar{\psi}\gamma^\mu\lambda^a\psi)^2 +  (\bar{\psi}\gamma^\mu\gamma_5\lambda^a\psi)^2 \Big],
\label{Lag_vec}
\end{align}
\end{widetext}
where $\lambda^a$ ($a=1,2...8$) are the Gell-Mann matrices of the SU(3) group and $\lambda^0=\sqrt{\frac{2}{3}}\mathbb{1}$.

The values of the vector-type couplings in Eq. (\ref{Lag_vec}) can be fixed by fitting the meson properties in the vacuum \cite{Klimt:1989pm}, however, we will adopt a different strategy. We start by taking three scenarios for $\mathscr{L}_{vec}$:
\begin{widetext}
\begin{equation}
\mathscr{L}_{vec}=
\begin{cases}
    G_V\sum\limits_{a=0}^8\Big[ (\bar{\psi}\gamma^\mu\lambda^a\psi)^2 +  (\bar{\psi}\gamma^\mu\gamma_5\lambda^a\psi)^2 \Big], \quad \text{with } G_\omega=G_\rho=G_V & \mapsto \text{model NJL(V+P+VI+PI)} 
    \\
    G_V\Big[ (\bar{\psi}\gamma^\mu\lambda^0\psi)^2 + (\bar{\psi}\gamma_5\gamma^\mu\lambda^0\psi)^2 \Big],  \qquad \;\;\; \text{with } G_\rho=0; \,\, G_\omega=G_V  & \mapsto \text{ model NJL(V+P)} 
     \\
     G_V\sum\limits_{a=1}^8\Big[ (\bar{\psi}\gamma^\mu\lambda^a\psi)^2 +  (\bar{\psi}\gamma^\mu\gamma_5\lambda^a\psi)^2 \Big],  \quad \text{with } G_\omega=0; \,\, G_\rho=G_V  & \mapsto \text{model NJL(VI+PI)} 
\end{cases}.
\label{L_vect}
\end{equation}
\end{widetext}
We will take the ratio $\xi=G_V /G_S$ as a free parameter, with $G_S$ fixed as usual in the NJL-type models. As pointed out in \cite{Fukushima:2008wg}, there is still no constraint on $G_V$ at finite density, even if there are attempts in that direction \cite{Bratovic:2012qs}. 
Having no definitive knowledge not even on its sign, $G_V$ can be seen as describing effects induced in dense quark matter and might be related to an in-medium modification \cite{Fukushima:2008wg}.

We can also argue that the couplings $G_S$ and $G_D$ are not well constrained in the medium either, but we follow the usual strategy and fix their values to the vacuum  meson properties and take these values for all densities (and/or temperatures)\footnote{For example, the study on how the influence of the density in $G_D$ affects mesons properties were made in Ref. \cite{Costa:2005cz}.
}.

In model NJL(V+P+VI+PI) we take for the vector-type couplings the particular choice $G_\omega = G_\rho \equiv G_V$ (independently of the value of $G_V$ this choice makes the $\omega$ and $\rho$ mesons degenerate in the vacuum \cite{Lutz:1992dv}). 

The thermodynamic potential density (subtracting the zero-point energy contribution $\Omega_0$) for $G_\omega\ne G_\rho$ is 
\begin{widetext}
\begin{align}
\Omega-\Omega_0 &= 2G_S  \left(  \sigma_u^2 + \sigma_d^2 + \sigma_s^2 \right) -4G_D\sigma_u\sigma_d \sigma_s-\frac{2}{3}G_\omega\left( \rho_u + \rho_d + \rho_s \right)^2 -G_\rho\left( \rho_u - \rho_d  \right)^2 -\frac{1}{3}G_\rho\left( \rho_u + \rho_d - 2\rho_s \right)^2\nonumber\\  
& - 2 N_c\int \frac{d^3p}{(2\pi)^3} \sum_{i=u,d,s}  \Big[  E_i + T \ln \big( 1 + e^{-(E_i+\tilde{\mu}_i)/T}  \big) + T \ln \big( 1 + e^{-(E_i-\tilde{\mu}_i)/T}  \big) \Big],
\label{potSU3}
\end{align}
where $\sigma_i$ is the $i$-quark flavor condensate, $\rho_i$ is the $i$-quark flavor density (both presented in the Appendix \ref{Appendix}.) and $E_i=\sqrt{p^2+M_i^2}$. 
The effective chemical potentials for the quarks in the general case are given by\footnote{ The full expressions for each case are given in Appendix \ref{Appendix}.}:
\begin{eqnarray}
\tilde{\mu}_i&=&\mu_i-\frac{4}{3}\left[(G_\omega+2G_\rho)\rho_i+(G_\omega-G_\rho)\rho_j+(G_\omega-G_\rho)\rho_k\right], \,\,\,\,i\ne j\ne k\in \{u,d,s\}.
\end{eqnarray}
\end{widetext}

In the mean field approximation, we obtain the following \textit{gap} equations:
\begin{align}
 M_i-m_i&=-4G_S\sigma_i+2G_D\sigma_j \sigma_k,\\
&i\ne j\ne k\in \{u,d,s\}.\nonumber
\end{align}


To understand the role of strangeness in neutron stars we will also adopt a SU(2) NJL model with vector interaction (see for example Ref. \cite{Buballa:2003qv}).
Again, we study three cases for vector interactions. In SU(2) they are obtained from Eq. (\ref{Lag_vec}) by substituting the Gell-Mann matrices, $\lambda^a$, by the SU(2) Pauli matrices $\boldsymbol{\tau}$ matrices that act in flavor space (with $\tau^0 = \mathbb{1}$).

The thermodynamic potential density (subtracting the zero-point energy contribution $\Omega_0$) is now given by 
\begin{widetext}
\begin{align}
\Omega-\Omega_0 &= G_S\left( \sigma_u + \sigma_d \right)^2 - G_\omega\left( \rho_u + \rho_d \right)^2 - G_\rho\left( \rho_u - \rho_d \right)^2\nonumber\\
&-2 N_c\int \frac{d^3p}{(2\pi)^3} \sum\limits_{i=u,d}  \Big[  E_i + T \ln \big( 1 + e^{-(E_i+\tilde{\mu}_i)/T}  \big) + T \ln \big( 1 + e^{-(E_i-\tilde{\mu}_i)/T}  \big) \Big].
\label{potSU2}
\end{align}
\end{widetext}
The replacement of the Gell-Mann matrices in SU(3) by the Pauli matrices in SU(2) has a direct effect on the effective chemical potentials.
Indeed, for the general case ($G_\rho\neq G_\omega$) they became\footnote{The full expressions for each case in SU(2) are given in Appendix \ref{Appendix}.}:
\begin{align}
     \tilde{\mu}_i=\mu_i-2G_\omega\left( \rho_i + \rho_j \right)-4t_i G_\rho\left( \rho_i - \rho_j \right),
\end{align}
where $t_i$ is the isospin projection and takes the value +1/2 for the $u$-quark.
Finally, the \textit{gap} equations are,
\begin{equation}
M_i-m_i=-4G_S\left( \sigma_i +\sigma_j \right),\;\;i\neq j  \in \{ u,d\}.
\label{gaps}
\end{equation}

In the limit $T\rightarrow 0$ matter inside neutron stars is degenerate.
The pressure,
and the energy density,
\begin{equation}
\epsilon=\Omega+\sum\limits_{i=u,d}\mu_i\rho_i, 
\end{equation}
are given in Appendix \ref{Appendix}.

\subsubsection{Parameters of the models}
\label{Parameters}

In the SU$(2)$ NJL model, when equal current masses for each quark flavor are considered, there are three free parameters: the current quark mass $m_u=m_d=m$, the coupling $G_S$, and the cutoff, $\Lambda$, that regularizes the model. Indeed, the NJL model is not renormalizable and there are different ways to regularize the model (see for example \cite{Moreira:2010bx}). In this work, we will consider a sharp cutoff, $\Lambda$, in 3-momentum space.

The parameters of the model are fixed in order to reproduce the experimental values for the mass and decay constant of the pion ($m_{\pi}=135.0$ MeV and $f_{\pi}=92.4$ MeV) and the value of the quark condensate in the vacuum.

\begin{table}[H]
\begin{ruledtabular}
    \begin{tabular}{lccccc}
    set  & $\Lambda$ & $m_{u,d}$ & $G_S\Lambda^2\,\,\,\,\,\,\,\,$ &  $-\braket{\bar{u}u}^{1/3}$ & $M_{u,d}$ \\
     &  [MeV] & [MeV] & & [MeV] &[MeV] \\
    \hline
    SU(2) & 648.0 & 5.1   & 2.110 & 248.2  & 312.6 \\
    \end{tabular}
\end{ruledtabular}
   \caption{Sets of parameters used throughout the work and reproduced observables in the vacuum, for each parametrization. $\Lambda$ is the model cutoff, $m_{u,d}$ is the quark current mass, and $G_S$ is the coupling constant. The results for the \textit{u}-quark condensate, $\braket{\bar{u}u}$, and for the constituent masses, $M_{u,d}$, are also presented.}
  \label{tab:1}
\end{table}

Since we are interested in studying hybrid neutron stars containing a hadronic  and a quark phase,  a NJL model parametrization that reproduces in the vacuum the same baryonic chemical potential as the hadronic model should be considered, i.e. a parametrization that gives, in the vacuum, $M_u=M_d\approx 313$ MeV, about one third of the vacuum nucleon mass.
We propose the new set of parameters for the SU(2) model,  see Table \ref{tab:1}, that gives $m_{\pi}=135$ MeV,  $f_{\pi}=92.4$ MeV and $\braket{\bar{u}u} = (-248.2$ MeV$)^3$. 

As already mentioned, the parameter $G_V$ in the vectorial terms is seen as a ``free'' parameter and consequently, in the present work we study several values of the ratio $\xi=G_V/G_S$.

In the $T=0$ limit for stellar matter application, we define the ratio between the Fermi's moment for each flavor of quark $(\lambda_{F_i})$, and the model's cutoff ($\Lambda$) as the limit of applicability of our model: the model is valid for densities and/or chemical potentials that verify $\lambda_{F_i}/\Lambda \leq 1 $. In SU$(2)$, the studied models are still valid at about $\rho_B \approx 11\rho_0$ (where $\rho_0=0.16$ fm$^{-3}$ is the saturation density), a far larger density than the ones found inside neutron stars.

As previously in the SU$(2)$ case, we propose a new parametrization for the SU(3) case which reproduces the same baryonic chemical potential at zero density  in both quark and hadronic phases (implying that $M_u=M_d\approx 313$ MeV).
This new parametrization is presented in Table \ref{tab:2}.  In Table \ref{tab:3}, we compare the values of the calculated observables with the respective experimental values. 

\begin{table}[t]
\begin{ruledtabular}
    \begin{tabular}{lccccccc}
   set   & $\Lambda$ & $m_{u,d}$& $m_s$ & $G_S\Lambda^2\,\,\,$ & $G_D\Lambda^5\,\,\,\,\,$ & $M_{u,d}$ & $M_s$ \\
& [MeV] & [MeV] & [MeV] & & & [MeV] & [MeV] \\
    \hline
    SU(3) & 630.0 & 5.5   & 135.7 & 1.781 &  9.29 & 312.2   & 508 \\
    \end{tabular}
\end{ruledtabular}
  \caption{$\Lambda$ is the model cutoff, $m_{u,d}$ and $m_{s}$ are the quark current masses, $G_S$ and $G_D$ are coupling constants. $M_{u,d}$ and $M_{s}$ are the resulting constituent quark masses in the vacuum.}
  \label{tab:2}
\end{table}

As in the SU(2) case, we restrict the applicability of the models in SU(3), in the $T\rightarrow 0$ limit, to the density at which the ratio $\lambda_{F_i}/\Lambda \leq 1$. The models in SU$(3)$ are valid until at least $15\rho_0$, densities well above those found inside neutron stars.

\begin{table}[t!]
\begin{ruledtabular}
    \begin{tabular}{ccc}
     & SU(3) {   }& Experimental \cite{Agashe:2014kda} \\
        \hline
    $m_{\pi^{\pm}}$ [MeV]     & 138.5  & 139.6 \\
    $f_{\pi^{\pm}}$ [MeV]     & 90.7   & 92.2 \\
    $m_{K^{\pm}}$ [MeV]       & 493.5  & 493.7 \\
    $f_{K^{\pm}}$ [MeV]       & 96.3   & 110.4 \\
    $m_{\eta}$ [MeV]    & 478.2  & 547.9 \\
    $m_{\eta'}$ [MeV]   & 953.7  & 957.8 \\
    \end{tabular}
\end{ruledtabular}    
    \caption{Masses and decay constants of several mesons within the model and the respective experimental values.}
  \label{tab:3}
\end{table}

\subsection{$\beta$-equilibrium matter}
\label{beta_eq}
In order to study cold stellar matter, $\beta$-equilibrium and charge neutral matter must be imposed and, therefore, a leptonic contribution must be added to the Lagrangian densities (\ref{Lag_hadron}) and (\ref{Lag_quark}),
\begin{eqnarray}
\mathscr{L}_l=\sum_{l=e,\mu}\bar{\psi}_l(i\slashed{\partial}+m_l)\psi_l.
\end{eqnarray}
The leptonic contribution to thermodynamic potential densities of the models considered is
\begin{align}
\Omega_l &= 2T \sum_{l=e,\mu}\int \frac{d^3p}{(2\pi)^3}  \Big[ \ln \big( 1 + e^{-(E_l+\mu_l)/T}  \big) \nonumber\\
&+ \ln \big( 1 + e^{-(E_l-\mu_l)/T}  \big) \Big],
\label{potSU2_e}
\end{align}
where $E_l=\sqrt{p^2+m_l^2}$, and the sum is over electrons and muons.
At $T=0$ the mean free path of  neutrinos is larger than the star radius and we will consider that they escape and that they have a zero chemical potential.

Neutrality and $\beta$-equilibrium for the hadronic matter results in the conditions
\begin{align}
\rho_p = \rho_e+\rho_\mu.
\end{align}
and
\begin{align}
\mu_n-\mu_p = \mu_e,
\end{align}
The corresponding conditions for quark matter read
\begin{equation}
\frac{1}{3}\left( 2\rho_u-\rho_d-\rho_s \right) -\rho_e -\rho_\mu=0\\
\end{equation}
and, 
\begin{equation}
\mu_d=\mu_s=\mu_u+\mu_{e}. 
\end{equation}
In the SU(2) NJL model, $s$-quarks are not present and, therefore,  $\rho_s$=0.

All thermodynamic quantities of interest, e.g. the pressure and the energy density are presented in the Appendix \ref{Appendix} (in the limit $T\rightarrow 0$).


\subsection{Phenomenological bag constant and Gibbs construction}
\label{BAG}

As pointed out in Ref. \cite{Pagliara:2007ph} the pressure within the NJL-type models is defined up to a constant $B$, similar to the MIT bag constant. 
This constant is usually fixed by requiring that the corrected pressure $P$ goes to zero at vanishing baryonic chemical potential (a detailed study of the bag pressure in NJL model was done in Ref. \cite{Menezes:2003xa}).

However, the procedure used to fix the effective bag constant within NJL models is crucial for the stability of the star when the phase transition to quark matter is considered.
In the same work \cite{Pagliara:2007ph},  the  bag constant $B^*$ is introduced and is fixed imposing that the deconfinement occurs at the same baryonic chemical potential, $\mu_B^{\text{crit}}$, as the chiral phase transition.
In the present work we consider the  NL3$\omega\rho$ model   (see Sec. \ref{hadronicEOS}) to describe the hadronic phase and compute the transition to quark matter imposing Gibbs conditions and  the coincidence between the  deconfinement phase transition  and  the partial restoration  of the  chiral symmetry.  This is achieved by adding to the quark EoS [Eq. (\ref{potSU3}) in SU(3) and Eq. (\ref{potSU2}) in SU(2)] the suitable value of the bag constant, $B^*$. For comparison we  will also study the  $B^*=0$ case.
Including $B^*$ modifies  the  quark matter EoS  in the following way:
\begin{equation}
P_{eff}=P+B^*,\quad \epsilon_{eff}=\epsilon-B^*,
\end{equation}
and, therefore, shifts the pressure to larger values for a given baryonic chemical potential, favoring the hadron-quark phase transition.

To build the hybrid EoS we use the Gibbs conditions: both phases must be in chemical, thermal and mechanical equilibrium
\begin{equation}
\mu_B^H= \mu_B^Q   \;\;\;\wedge\;\;\;  p_B^H= p_B^Q  \;\;\;\wedge\;\;\; T_B^H= T_B^Q =0,
\end{equation}
where the $H$ and $Q$ indices represent, respectively, the confined (hadronic) and deconfined (quark) phases.

The chiral symmetry transition point ($\mu_B^{\text{crit}}$) is defined in the following way: if the phase transition is of first-order, we search for the $\mu_B$ at which there is a discontinuity in the quark condensate (the order parameter): 
the stable solutions of the gap equations are realized by the minimum of the thermodynamic potential or, equivalently, maximum of the pressure (see Ref. \cite{Costa:2010zw} for details).
If the transition is a crossover, we search for the zeros of the second derivative of the light quark condensates, $\partial^2 \braket{\bar{q}_i q_i}/\partial{\mu_B}^2=0$.  In the cases where there are different chemical potentials for each quark flavor (different phase transitions for each flavor), the chemical potential used in the Gibbs condition is given by the average of the baryonic chemical potentials at the corresponding phase transitions.
\begin{equation}
\mu^{\text{crit}}_B=\frac{\mu_{B(u)}^{\text{crit}}+\mu_{B(d)}^{\text{crit}}}{2}.
\end{equation}

\begin{table}[t]
\begin{ruledtabular}
    \begin{tabular}{cccc}
   Model (SU(2)) & $\xi$ & Type  & $\mu_B^{\text{crit}}$ [MeV]\\
    \hline
      NJL                                 & 0.00  & 1st-order   & 1119\\ \cline{2-4}
    \multirow{2}[6]{*}{NJL(V+P+VI+PI)}    & 0.25  & crossover   & 1055\\
                                          & 0.50  & crossover   & 1099\\
                                          & 0.75  & crossover   & 1149\\ \cline{2-4}              
    \multirow{2}[6]{*}{NJL(V+P)}          & 0.25  & crossover   & 1051\\
                                          & 0.50  & crossover   & 1089\\
                                          & 0.75  & crossover   & 1134\\ \cline{2-4}
    \multirow{2}[6]{*}{NJL(VI+PI)}        & 0.25  & crossover   & 1022\\
                                          & 0.50  & crossover   & 1025\\
                                          & 0.75  & crossover   & 1029\\
    \end{tabular}
\end{ruledtabular}    
  \caption{Type of the chiral symmetry phase transition and respective baryonic chemical potential ($\mu_B^{\text{crit}}$), for each value of $\xi$, model and parameter set.}
  \label{tab:4}
\end{table}%


\section{Results and discussion}
\label{results}

In the present section we present our results and discuss the possible existence of hybrid
stars within the NJL model, for the three scenarios previously defined.
The neutron star mass and radius are obtained solving the Tolmann-Oppenheimer-Volkov (TOV) equations \cite{tov,Tolman:1939jz}. In particular, for each  star we calculate  the maximum gravitational mass and the respective central density, radius and  maximum baryonic mass.
We also investigate the role of strangeness in the EoS.
For each case we consider $\xi=G_V/G_S=$ 0, 0.25, 0.5, and 0.75, with $G_S$ fixed.

\subsection{Results without strangeness}
\label{res_SU(2)}

\begin{figure*}[t!]
\begin{tabular}{cc}
\includegraphics[width=0.43\textwidth]{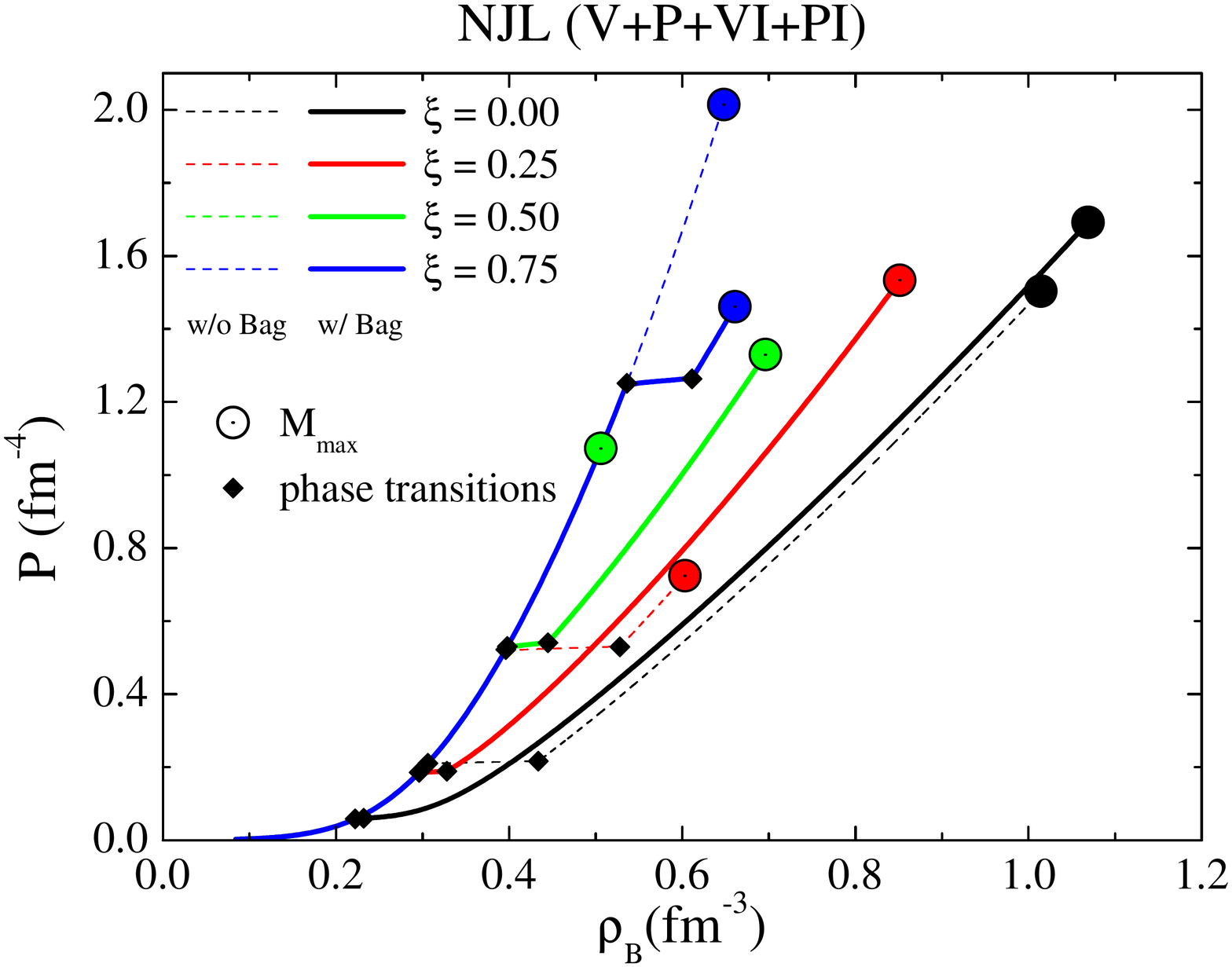}\put(-30,30){\textbf{(a)}}&
\includegraphics[width=0.43\textwidth]{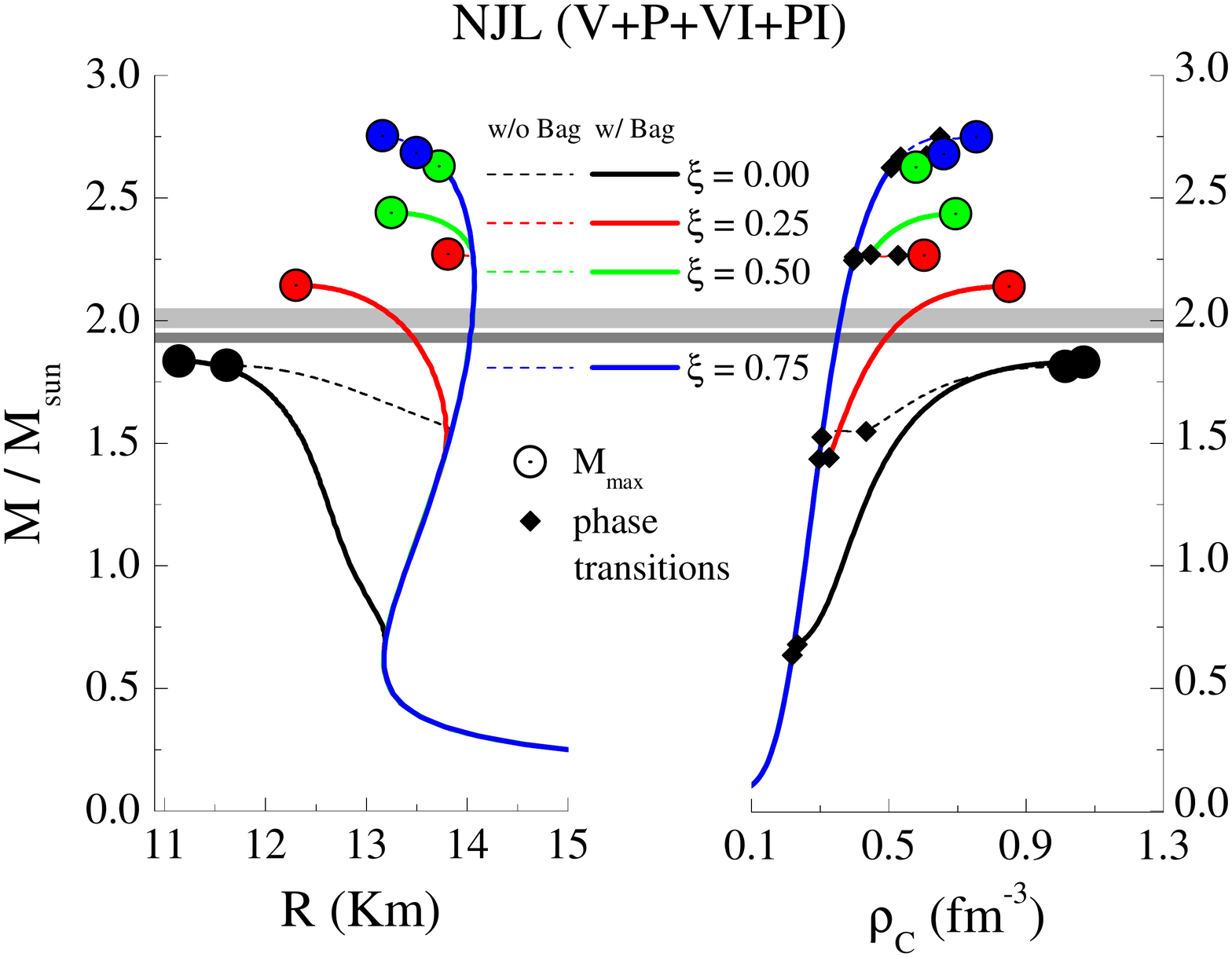}\put(-35,35){\textbf{(d)}}\\
\includegraphics[width=0.43\textwidth]{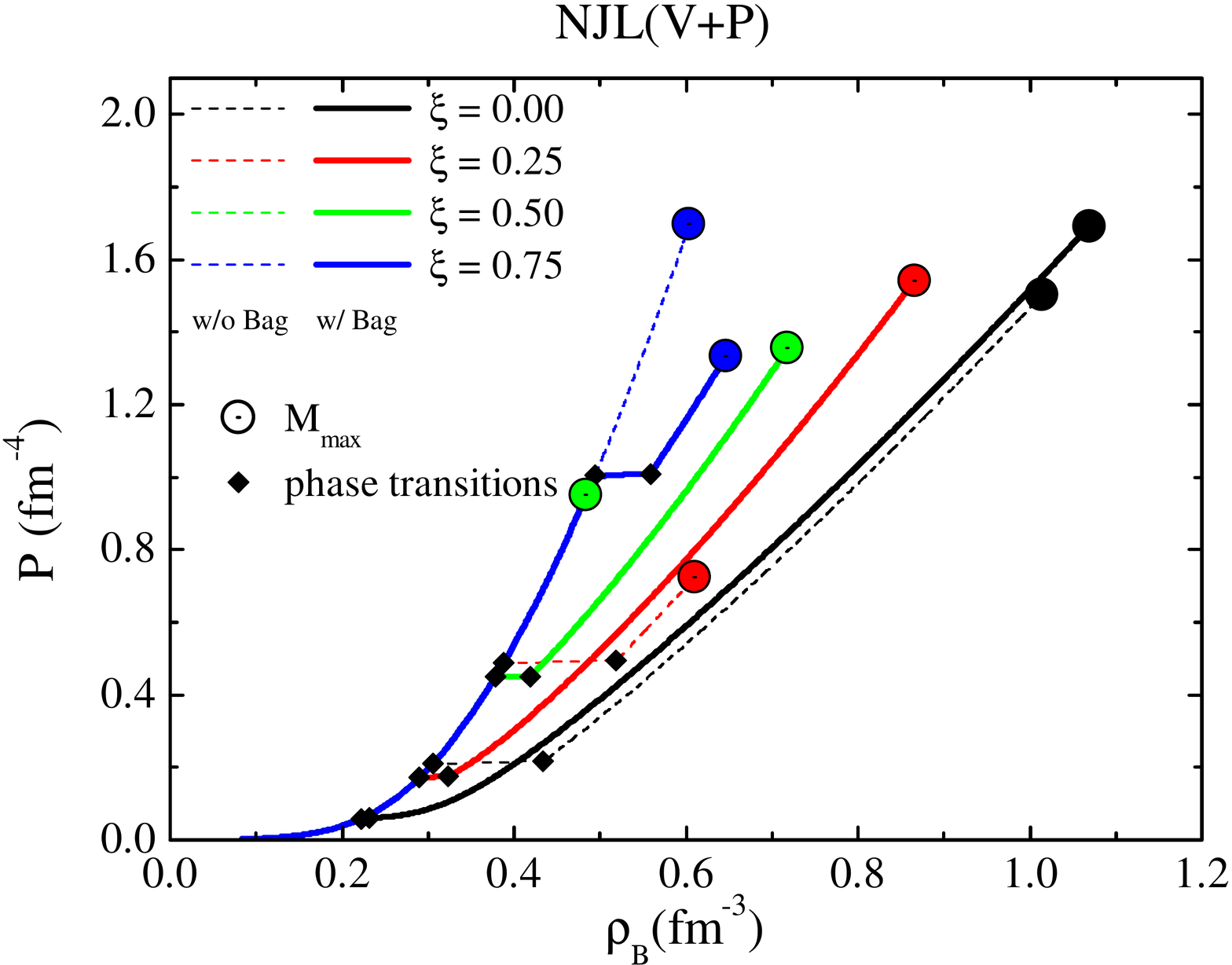}\put(-30,30){\textbf{(b)}}&
\includegraphics[width=0.43\textwidth]{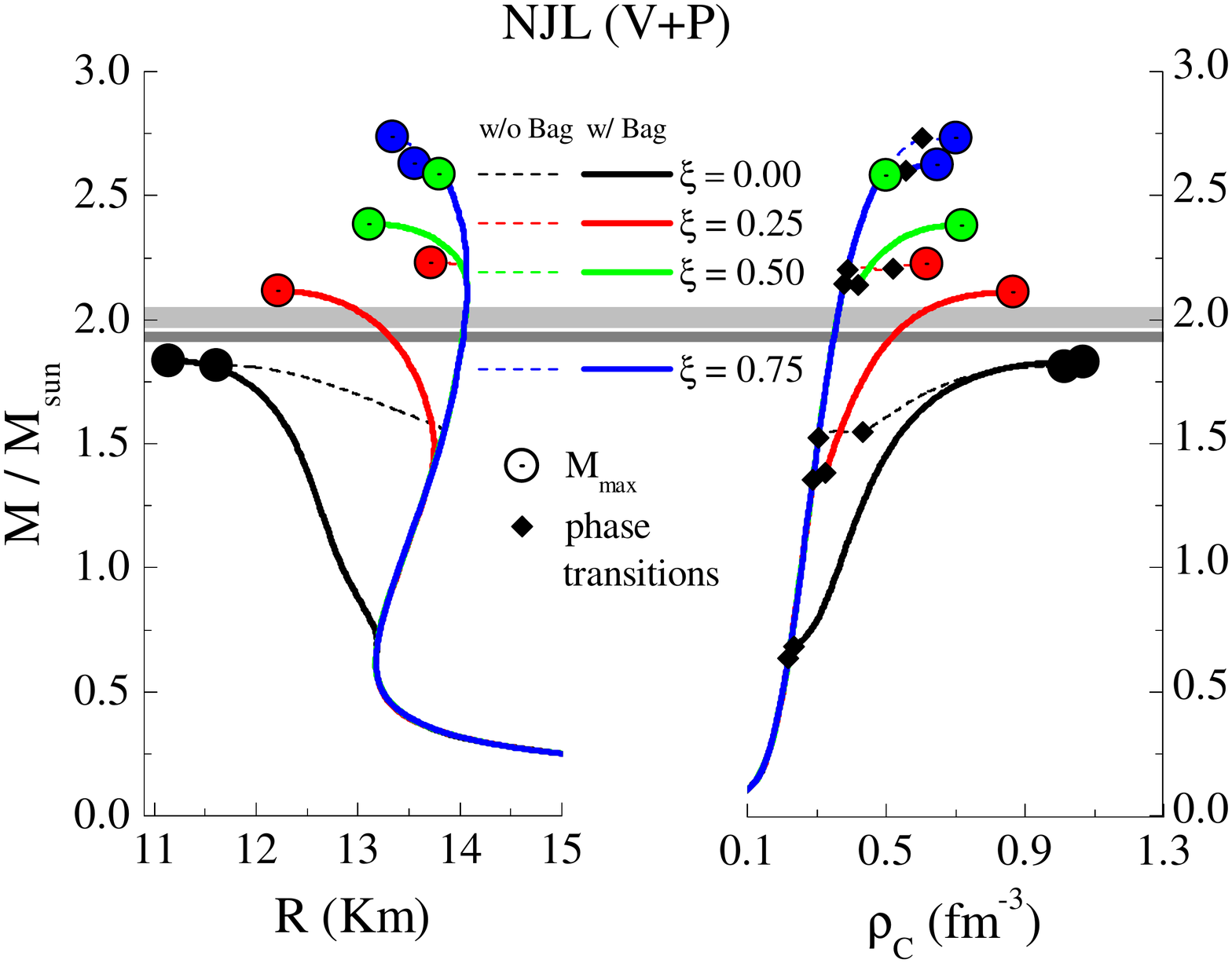}\put(-35,35){\textbf{(e)}}\\
\includegraphics[width=0.43\textwidth]{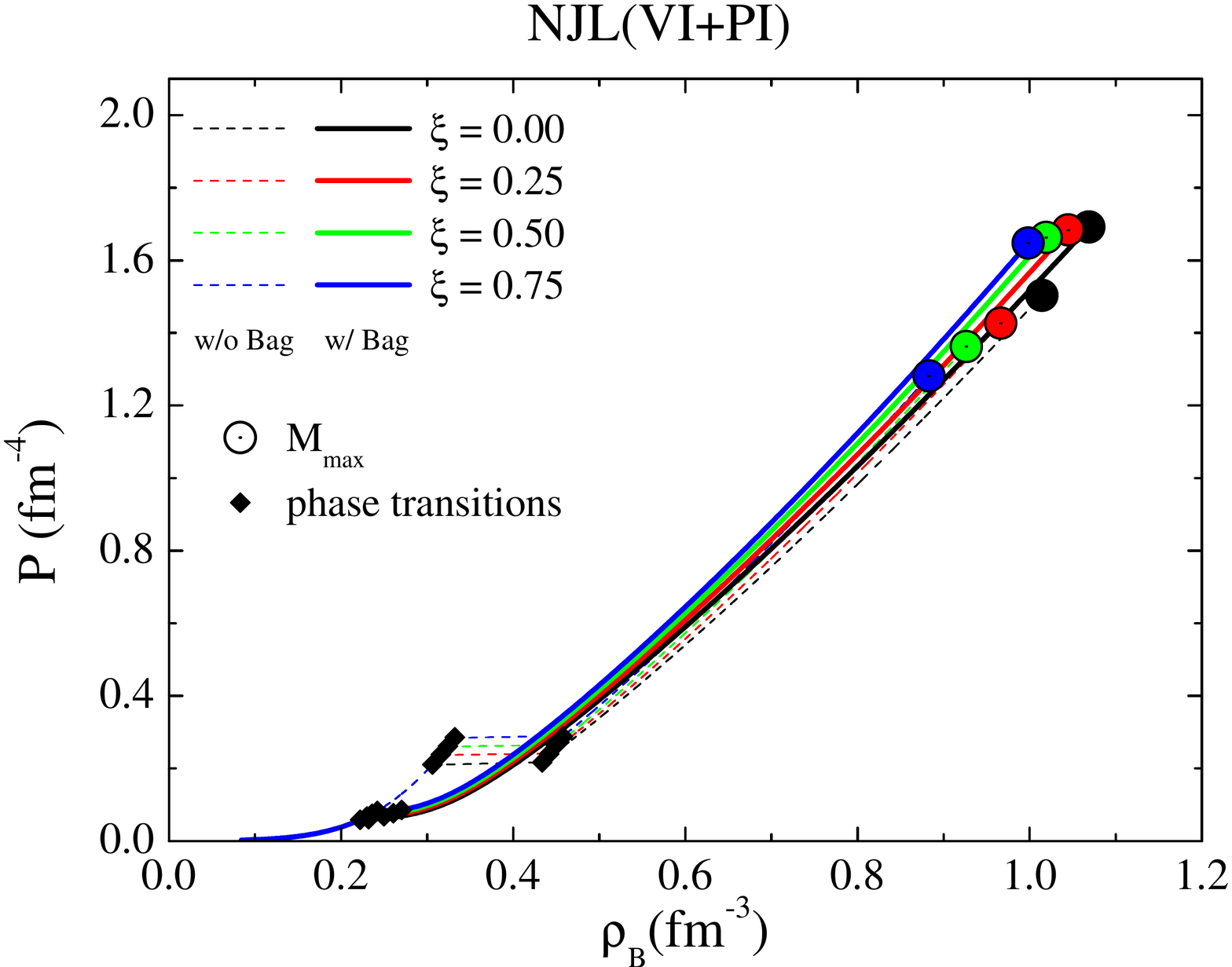}\put(-30,30){\textbf{(c)}}&
\includegraphics[width=0.43\textwidth]{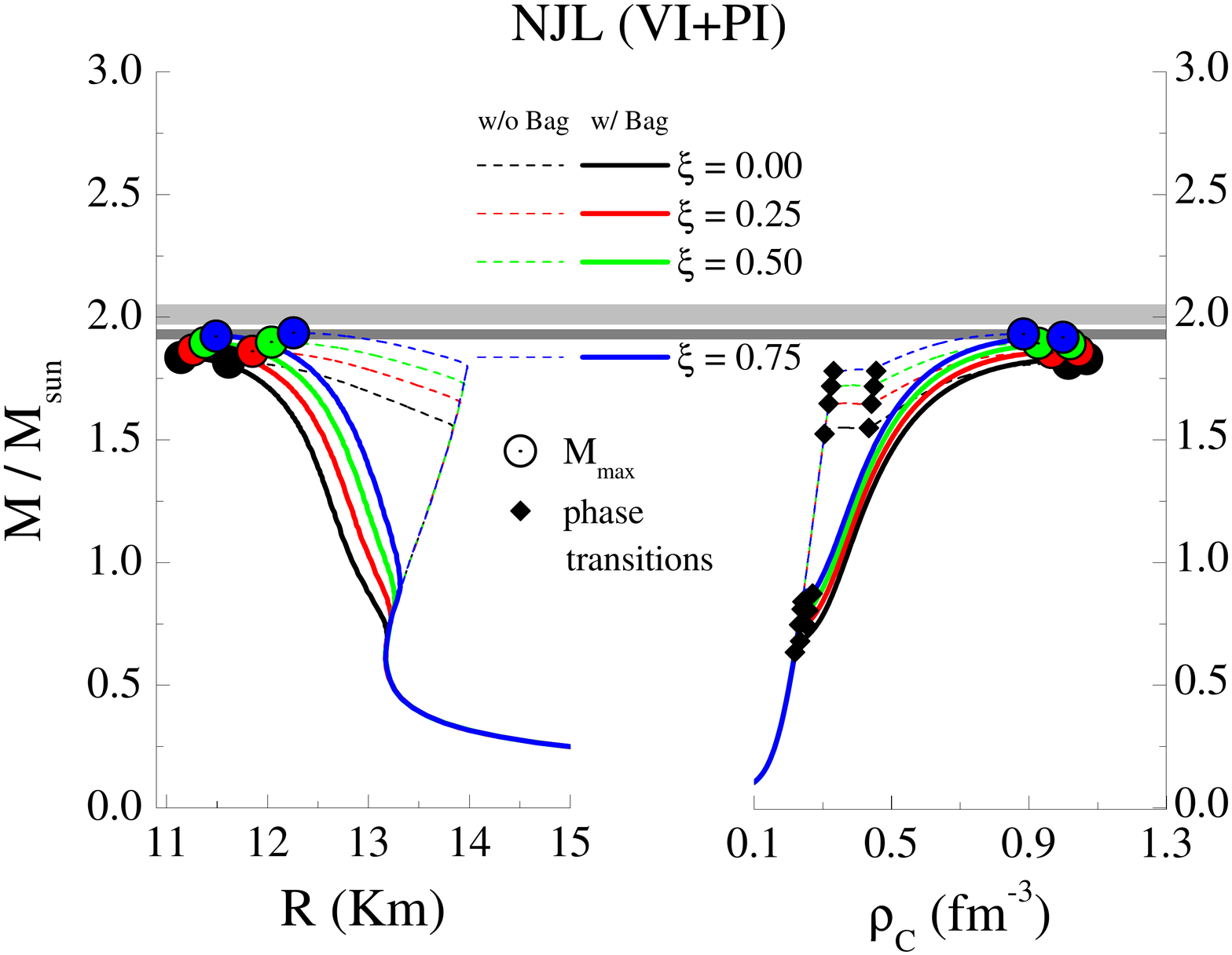}\put(-35,35){\textbf{(f)}}\\
\end{tabular}
\caption{{\it Left panels}: EoS for several values of $\xi$, for the SU(2) models of  NJL(V+P+VI+PI) [panel (a)], NJL(V+P) [panel (b)] and  NJL(VI+PI) [panel (c)] models. The star maximum mass, central density and confinement-deconfinement phase transitions are highlighted.
{\it Right panels}: mass-radius and mass-central density diagrams for several values of $\xi$ for the SU(2) NJL(V+P+VI+PI) [panel (d)], NJL(V+P) [panel (e)] and  NJL(VI+PI) [panel (f)] models. The star maximum mass, central density and confinement-deconfinement phase transitions are highlighted. 
The light-gray bar represents the mass constraint of the J0348+043 pulsar
($M = 2.01\pm0.04 M_\odot$) \cite{Antoniadis:2013pzd} 
 while the dark-gray bar the J1614-2230 pulsar ($M = 1.928\pm0.017 M_\odot$) \cite{Fonseca:2016tux}.
}
\label{fig:1}
\end{figure*}

\begin{table*}[t]
\begin{ruledtabular}
    \begin{tabular}{ccccccccccc}
    \multirow{2}{*}{Model} &  \multirow{2}{*}{$\xi$} & $B^*$ 
    & $\mu_B^{H-Q}$ & $\rho^H$  & $\rho^Q$  & $\rho^c$  & $M_m $  & $M_{bm}$  & $R_m $ & $R_{1.4} $ \\
    &     & [MeV$\,$fm$^{-3}$]
    & [MeV] & [fm$^{-3}$] & [fm$^{-3}$] & [fm$^{-3}$] & [M$_{\odot}$] & [M$_{\odot}$] & [km] & [km] \\
    \hline
     NJL & 0.00  & 0 & 1134 & 0.306 & 0.434 & 1.015 & 1.82  & 2.07  & 11.62 & 13.74 \\ \cline{2-11}
      & {\bf 0.11}  & {\bf}  & {\bf 1204} & {\bf 0.344} & {\bf 0.472} & {\bf 0.823} & {\bf 2.00} & {\bf 2.30} &  {\bf 12.56} & {\bf 13.74} \\ 
     NJL  & 0.25 & \multirow{2}{*}{0}  & 1308 & 0.396 & 0.528 & 0.603 & 2.27  & 2.67  & 13.81  & 13.74\\
     (V+P+VI+PI) & 0.50 &  & 1548 & 0.506 & 0.658 & 0.580 & 2.63 & 3.19  & 13.72 & 13.74\\
      & 0.75 &   & 1869 & 0.648 & 0.824 & 0.756 & 2.75  & 3.38  & 13.16 & 13.74\\ \cline{2-11}
    & {\bf 0.12}  & {\bf } & {\bf 1202} & {\bf 0.344} & {\bf 0.470} & {\bf 0.823} & {\bf 2.00} & {\bf 2.30} & {\bf 12.56} & {\bf 13.74}\\
    NJL   & 0.25 & \multirow{2}{*}{0}  & 1289 & 0.388 & 0.518 & 0.616 & 2.23  & 2.61  & 13.72  & 13.74\\
    (V+P) & 0.50 &  & 1497 & 0.484 & 0.630 & 0.501 & 2.58  & 3.12  & 13.80 & 13.74\\
          & 0.75 &   & 1769 & 0.604 & 0.771 & 0.700 & 2.74  & 3.36  & 13.34 & 13.74\\ \cline{2-11}
       & 0.25  &   & 1148 & 0.316 & 0.442 & 0.967 & 1.86  & 2.12  & 11.85 & 13.74\\
      NJL & 0.50 & \multirow{2}{*}{0} & 1163 & 0.324 & 0.450 & 0.928 & 1.90  & 2.17  & 12.04 & 13.74\\
      (VI+PI)   & 0.75 &   & 1177 & 0.332 & 0.458 & 0.884 & 1.94  & 2.22  & 12.26 & 13.74\\
                & {\bf 1.13}  & {\bf } & {\bf 1200} & {\bf 0.344} & {\bf 0.470} & {\bf 0.814} & {\bf 2.00} & {\bf 2.29} &  {\bf 12.61} & {\bf 13.74}\\                                
    \hline     
    NJL & 0.00  & 9.84 & 1020 & 0.222 & 0.232 & 1.068 & 1.84  & 2.11  & 11.14 & 12.48\\ \cline{2-11}
    & {\bf 0.13}  & {\bf 12.32} & {\bf 1063} & {\bf 0.260} & {\bf 0.293} & {\bf 0.948} & {\bf 2.00} & {\bf 2.31} & {\bf 11.77} & {\bf  13.74}\\ 
    NJL   & 0.25 & 15.16  & 1116 & 0.296 & 0.328 & 0.851 & 2.14  & 2.50  & 12.30  & 13.74\\
    (V+P+VI+PI) & 0.50 & 22.09 & 1313 & 0.398 & 0.445 & 0.695 & 2.44  & 2.91  & 13.25 & 13.74\\
      & 0.75 & 30.84  & 1616 & 0.536 & 0.611 & 0.660 & 2.69  & 3.27  & 13.50 & 13.74\\ \cline{2-11}
    & {\bf 0.15}  & {\bf 12.40} & {\bf 1067} & {\bf 0.264} & {\bf 0.298} & {\bf 0.941} & {\bf 2.00} & {\bf 2.32} & {\bf 11.80} & {\bf 13.74}\\ 
    NJL   & 0.25 & 14.50 & 1105 & 0.290 & 0.323 & 0.866 & 2.12  & 2.46  & 12.22 & 13.74\\
    (V+P) & 0.50 & 20.54 & 1268 & 0.378 & 0.419 & 0.718 & 2.39  & 2.83  & 13.11 & 13.74\\
          & 0.75 & 28.10 & 1519 & 0.494 & 0.558 & 0.647 & 2.63  & 3.19  & 13.55 & 13.74\\ \cline{2-11}
      & 0.25 & 10.29 & 1027 & 0.230 & 0.250 & 1.045 & 1.87  & 2.15  & 11.26 & 12.67\\
      NJL     & 0.50 & 10.75 & 1034 & 0.236 & 0.261 & 1.020 & 1.90  & 2.19  & 11.38 & 12.83\\
      (VI+PI) & 0.75 & 11.22 & 1041 & 0.242 & 0.270 & 0.999 & 1.92  & 2.22  & 11.49 & 12.99\\
              & {\bf 1.75}  & {\bf 13.13} & {\bf 1074} & {\bf 0.268} & {\bf 0.301} & {\bf 0.921} & {\bf 2.00} & {\bf 2.33} & {\bf 11.92} & {\bf  13.74}

    \end{tabular}
\end{ruledtabular}    
    \caption{Baryonic chemical potential ($\mu_B^{H-Q}$), hadron  ($\rho^H$) and quark ($\rho^Q$) baryonic density at  deconfinement and respective value of the parameter $B^*$. Values of central baryonic density ($\rho^c$), maximum gravitational mass ($M_m $), maximum baryonic mass ($M_{bm}$), radius ($R_m $), and radius of the $1.4M_\odot$ ($R_{1.4} $), for each model and  $\xi$ value, for the different models in SU(2). 
In bold we present the approximate values of $\xi$ at which $2M_\odot$ are obtained.
}
  \label{tab:6}
\end{table*}

We first study the SU(2) NJL case, which means that no strangeness is present in the system.
We recall that the parameters of the model have been determined so that 
in the vacuum the model has the same baryonic chemical potential as the hadronic
model. 

Table \ref{tab:4} shows the order of the chiral symmetry transition for different values of $\xi$, which were taken at $B^*=0$, but are independent of the bag constant. 
It can be seen that for the studied values of $\xi\ne0$, the chiral transition is a crossover instead of a first-order phase transition. 
Besides,  the transition occurs for smaller chemical potentials for $\xi\ne0$. 

Several $\beta$-equilibrium stellar matter EoS with nonzero $B^*$, taking into account the hadron-quark phase transition, are shown in Fig. \ref{fig:1} [panels (a), (b), and (c)], for the different vector contributions. 
These EoS will be used to determine compact star properties in the following discussion. The maximum mass star configuration determines the maximum central density attained in a star described within a given model. Therefore, in these plots the large colored circles indicate the central density of the maximum mass configuration  and we do not show the EoS above this density.
Small black diamonds indicate the hadron-quark phase transition. 
In each plot results for both $B^*=0$ and $B^*\ne 0$ are included. From the analysis of these figures some comments may be drawn:
a) the inclusion of $B^*\ne0$ shifts the deconfinement phase transition to smaller densities, allows the appearance of a quark phase even for a large value of $\xi$ and gives rise to larger central densities;
b) increasing the coupling $\xi$ in models  with vector-isoscalar terms makes the EoS harder  as shown previously, see \cite{Hanauske:2001nc,Klahn:2006iw,Pagliara:2007ph,Bonanno:2011ch}, and central densities of maximum mass configurations are smaller; 
c) the vector-isovector term [NJL(VI+PI)] has a much smaller effect than the vector-isoscalar term [NJL(V+P)], although qualitatively similar; 
d) the model labeled NJL(P+V+PI+VI) incorporates the effects of models NJL(P+V) and NJL(PI+VI) and, therefore, may give rise to larger central pressures [see panel (a) of Fig. \ref{fig:1}]; 
e) the harder the quark EoS the larger the deconfinement density, the effect being much stronger if the vector-isoscalar term is included;
f) the EoS which only includes the vector-isovector term originates smaller deconfinement densities and smaller density gaps between the hadronic and the quark density at deconfinement, i.e. a smaller mixed phase. Within this interaction larger central densities, larger quark fractions and smaller radii are attained;
g) for all cases, the vector-isoscalar interaction allows that the star reaches $2M_{\odot}$ if $\xi$ is large enough (the respective values are given in Table \ref{tab:6}).

We have calculated the mass and radius of hybrid stars integrating the TOV equations \cite{tov,Tolman:1939jz}. In Fig. \ref{fig:1} [panels (d), (e) and (f)], the mass versus  radius and mass versus central density  curves of the families of stars described by the EoS discussed above are plotted, respectively, in left and right side of each panel. We have considered the Baym-Pethick-Sutherland EoS \cite{bps} for the outer crust and for the inner crust the inner crust NL3$\omega\rho$ EoS that describes the pasta phases within a Thomas-Fermi approach \cite{Grill:2014aea} and links smoothly to the core NL3$\omega\rho$ EoS.

Some properties of  the hybrid stars, in particular of the maximum mass configurations are summarized in Table \ref{tab:6}. These properties include: the  bag constant $B^*$, the baryonic chemical potential at the transition $\mu_B^{H-Q}$, the central baryonic density $\rho^c$, the gravitational $M_{m}$ and baryonic mass $M_{bm}$ of the maximum mass configuration, and respective radius $R_m$, and the radius of the 1.4$M_\odot$ star.
 
The results show that even taking $B^*=0$ we have found stable hybrid stars with a pure quark core at the center ($\rho^c>\rho^Q$). All values of $\xi$ give rise stable hybrid stars if $B^*\ne0$,  but for $B^*=0$ stable hybrid stars are possible only  if the vector-isoscalar interaction is not too strong, (see Table \ref{tab:6}).

We verify that the vector-isoscalar has a very strong effect on the star structure giving rise to more massive stars, with larger radii and smaller quark contents, while the effect of the vector-isovector term on the maximum mass is very small (as it can be seen in Fig. \ref{fig:1} by comparing panels (d) and (e) with (f) ), and to get masses about $\sim 2\, M_\odot$ high values of $\xi$ ($\sim1.75$) are needed, see Table \ref{tab:6}. However, adding the vector-isoscalar interaction with a weak coupling would be enough to attain $M\gtrsim 2M_\odot$.


\subsection{The role of strangeness}
\label{res_SU(3)}

\begin{table}[t!]
\begin{ruledtabular}
    \begin{tabular}{cccc}
    Model (SU(3)) & $\xi$ & Type  & $\mu_B^{\text{crit}}$  [MeV] \\
    \hline
    NJL                                 & 0.00  & 1st-order  & 999 \\     \cline{2-4}
    \multirow{2}[7]{*}{NJL(V+P+VI+PI)}  & 0.25  & crossover  & 1023 \\
                                        & 0.50  & crossover  & 1052 \\
                                        & 0.75  & crossover  & 1087 \\     \cline{2-4}
    \multirow{2}[7]{*}{NJL(V+P)}        & 0.25  & crossover  & 1013 \\
                                        & 0.50  & crossover  & 1028 \\
                                        & 0.75  & crossover  & 1045 \\     \cline{2-4}
    \multirow{2}[7]{*}{NJL(VI+PI)}      & 0.25  & crossover  & 1008 \\
                                        & 0.50  & crossover  & 1018 \\
                                        & 0.75  & crossover  & 1028 \\
    \end{tabular}
\end{ruledtabular}    
  \caption{Type of the chiral symmetry phase transition and respective baryonic chemical potential ($\mu_B^{\text{crit}}$), for each value of $\xi$.}
  \label{tab:7}
\end{table}

\begin{figure*}[t!]
\begin{tabular}{cc}
\includegraphics[width=0.43\textwidth]{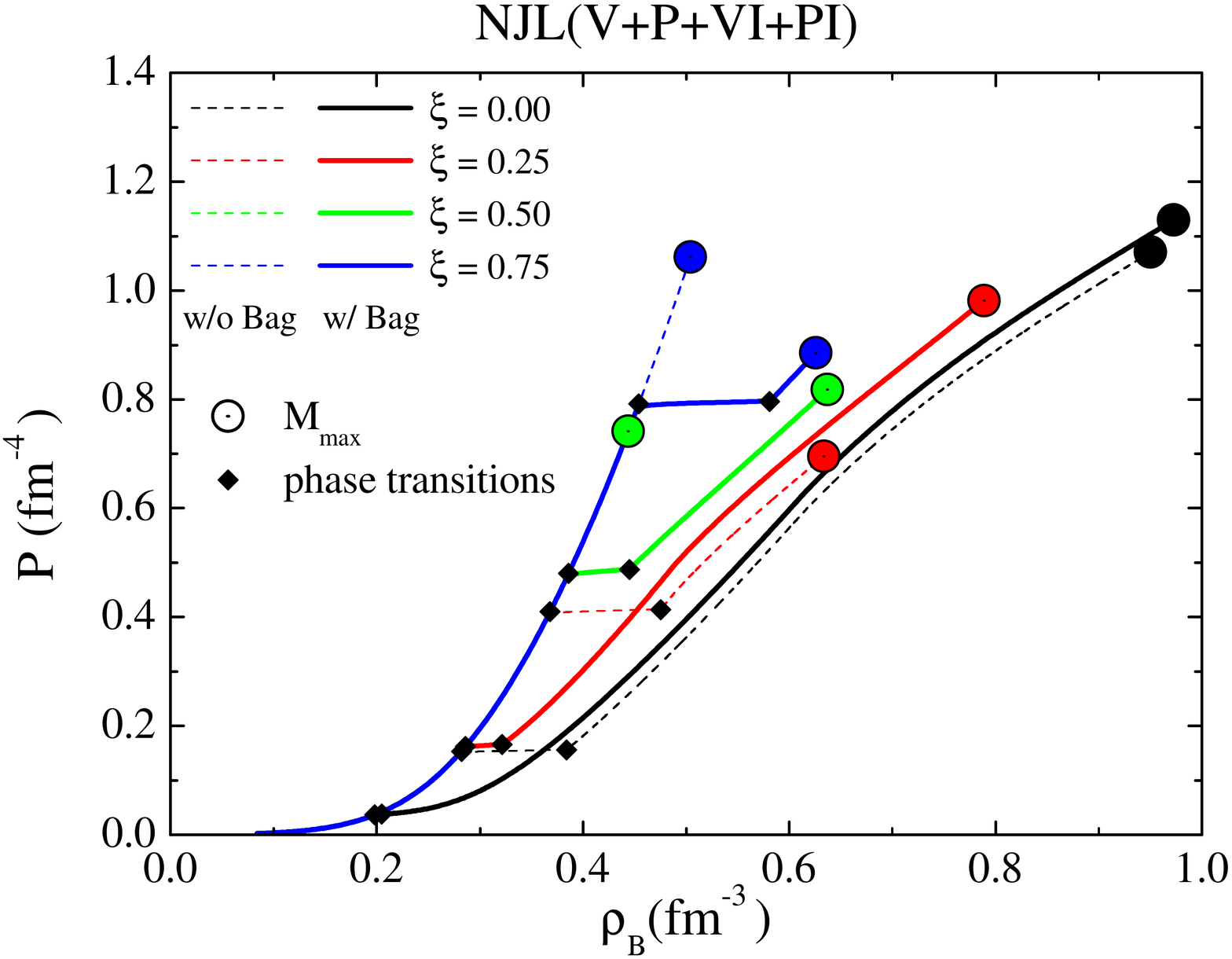}\put(-30,30){\textbf{(a)}}&
\includegraphics[width=0.43\textwidth]{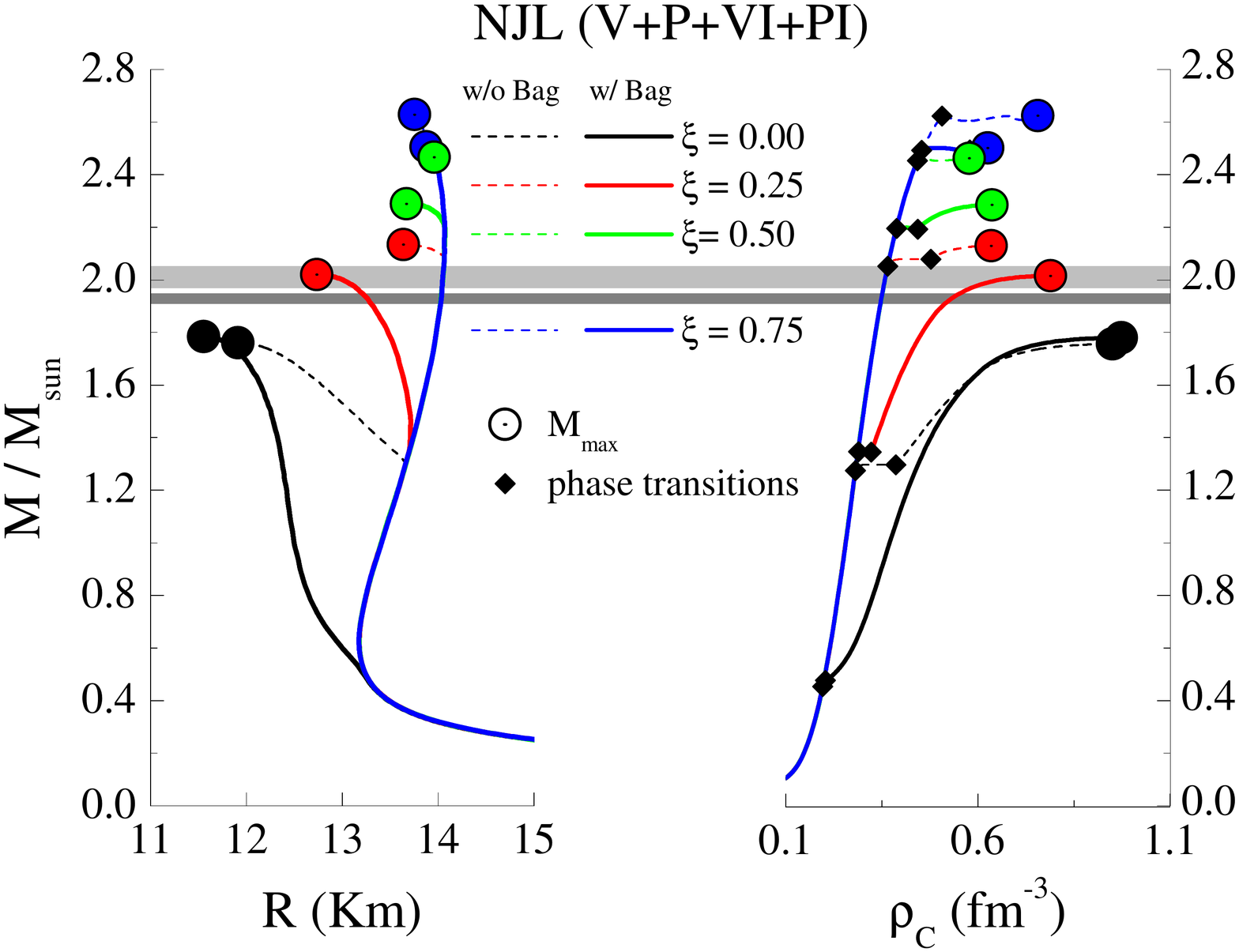}\put(-35,35){\textbf{(d)}}\\
\includegraphics[width=0.43\textwidth]{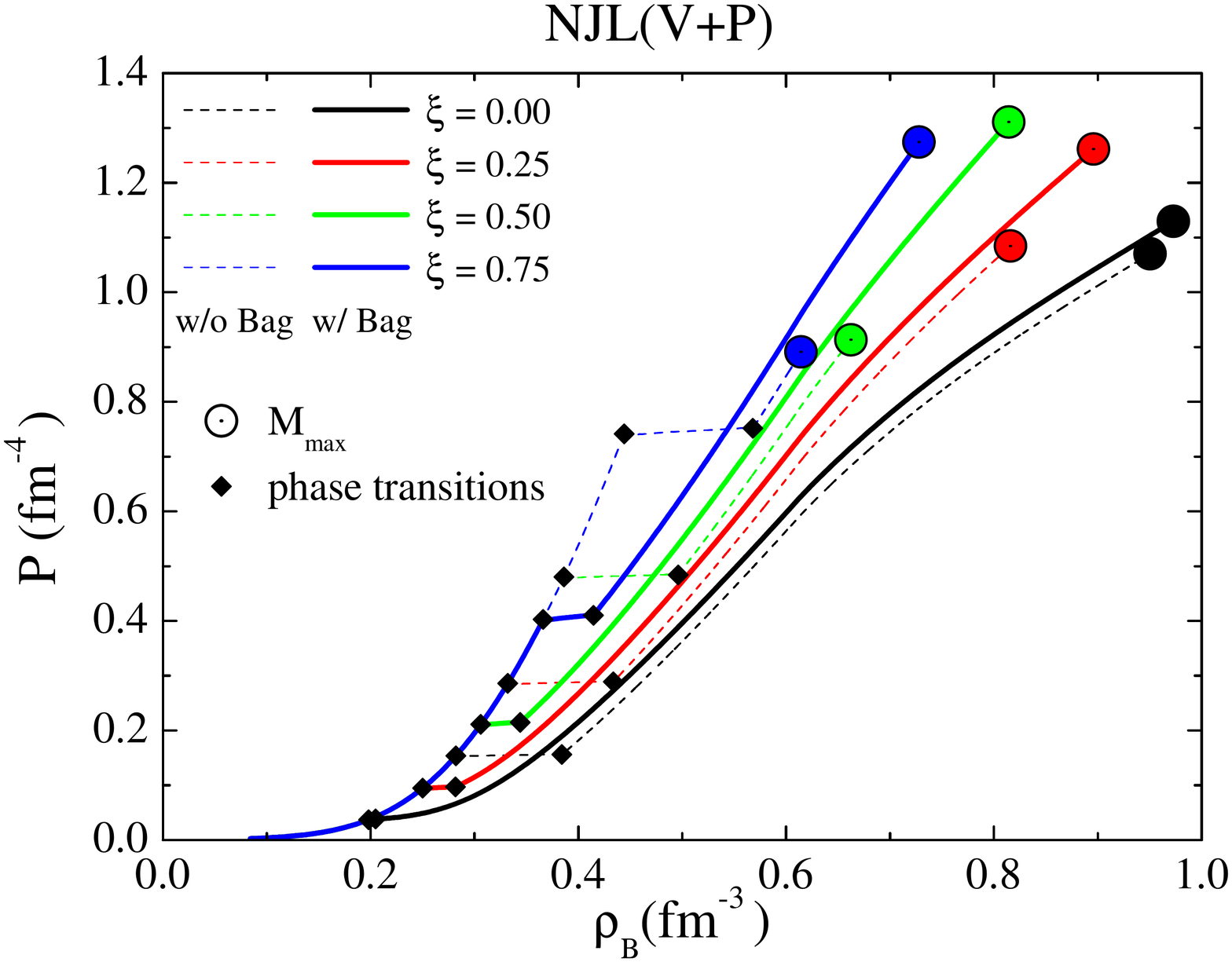}\put(-30,30){\textbf{(b)}}&
\includegraphics[width=0.43\textwidth]{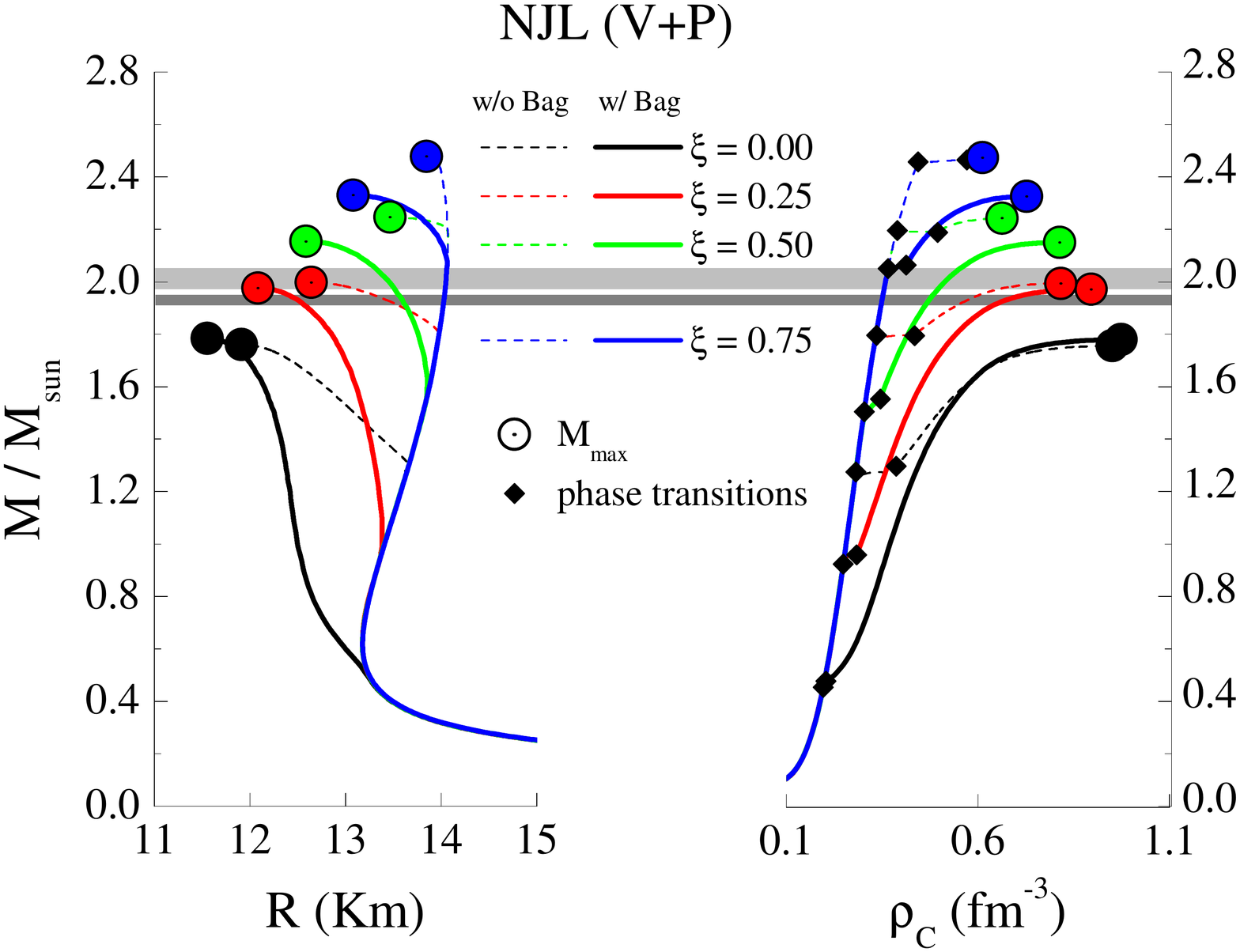}\put(-35,35){\textbf{(e)}}\\
\includegraphics[width=0.43\textwidth]{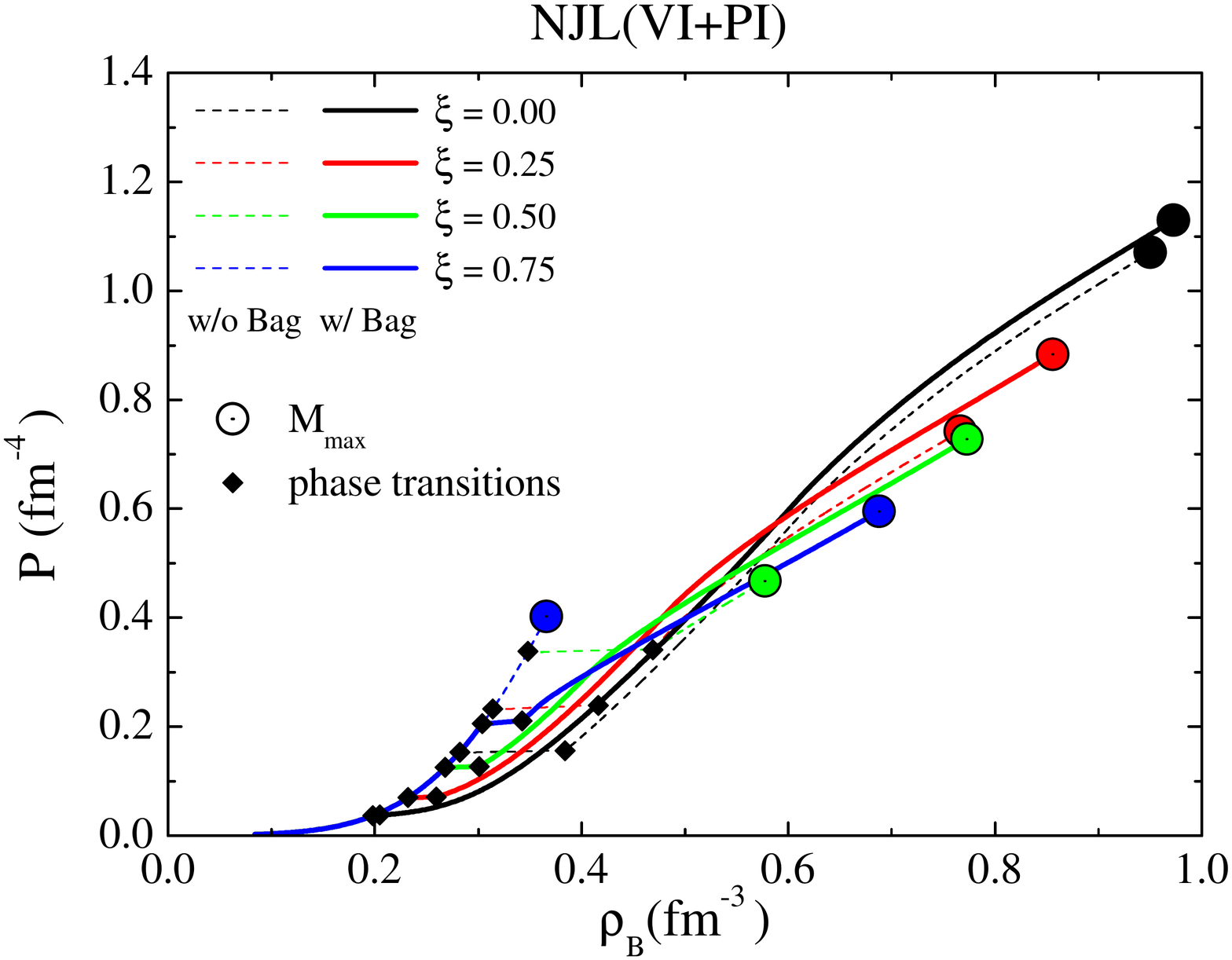}\put(-30,30){\textbf{(c)}}&
\includegraphics[width=0.43\textwidth]{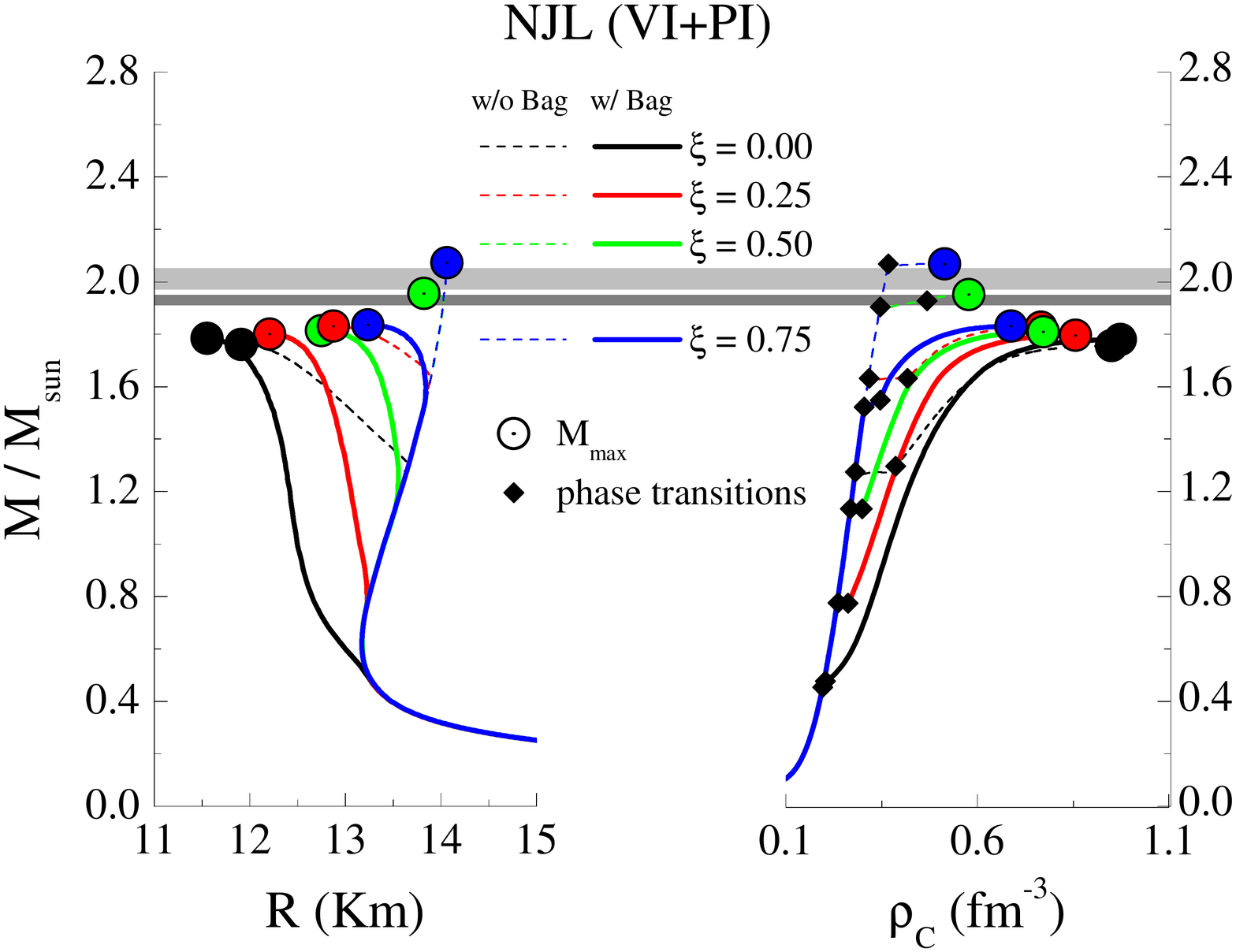}\put(-35,35){\textbf{(f)}}\\
\end{tabular}
\caption{{\it Left panels}: EoS for each value of $\xi$, for the NJL(V+P+VI+PI) [panel (a)], NJL(V+P) [panel (b)] and  NJL(VI+PI) [panel (c)] models. The star maximum mass, central density and confinement-deconfinement phase transitions are highlighted.
{\it Right panels}: mass-radius and mass-central density diagrams for each value of $\xi$ for the NJL(V+P+VI+PI) [panel (d)], NJL(V+P) [panel (e)] and  NJL(VI+PI) [panel (f)] models. The star maximum mass, central density and confinement-deconfinement phase transitions are highlighted.
The light-gray bar represents the mass constraint of the J0348+043 pulsar
($M = 2.01\pm0.04 M_\odot$) \cite{Antoniadis:2013pzd} 
 while the dark-gray bar the J1614-2230 pulsar ($M = 1.928\pm0.017 M_\odot$) \cite{Fonseca:2016tux}.
}
\label{fig:2}
\end{figure*}

In the previous section the strange degree of freedom was not considered, however it is expected that at large densities strangeness will set in. 
In this section we  take strangeness into account , and as before,  we will consider a parametrization that  predicts a vacuum constituent $u$ and $d$-quark mass equal to $\approx 313$ MeV, and that describes reasonably well the vacuum properties of several mesons,  see Table \ref{tab:2}.
All the features discussed in the previous section remain valid, as we may conclude analysing Table \ref{tab:7} where the type of phase transition is given for different strengths of the vector interaction, and  Fig. \ref{fig:2}  where the EoS [panels (a), (b) and (c)], 
and the  mass/radius and mass/density plots [panels (d), (e) and (f)] are presented.  The same conventions of Fig \ref{fig:1} are adopted.

The effect of $B^*$ and $\xi$ are the same as discussed in the previous section within the SU(2) NJL model. It should, however, be referred that care should be taken when comparing the SU(2) and SU(3) parametrizations: due to the different normalization of the Pauli and Gell-Mann matrices and the t' Hooft term.
Two solar mass stars are obtained  if the  vector-isoscalar interaction is strong enough, $\xi\gtrsim 0.17-0.28$ depending whether $B^*=0$ or $\ne0$, see Table \ref{tab:8}). 
Including only the vector-isovector interaction,  it is not possible to obtain a 2$M_\odot$ star with a quark core.

\begin{figure}[t]
\hspace{-0.2cm}
\includegraphics[width=0.40\textwidth]{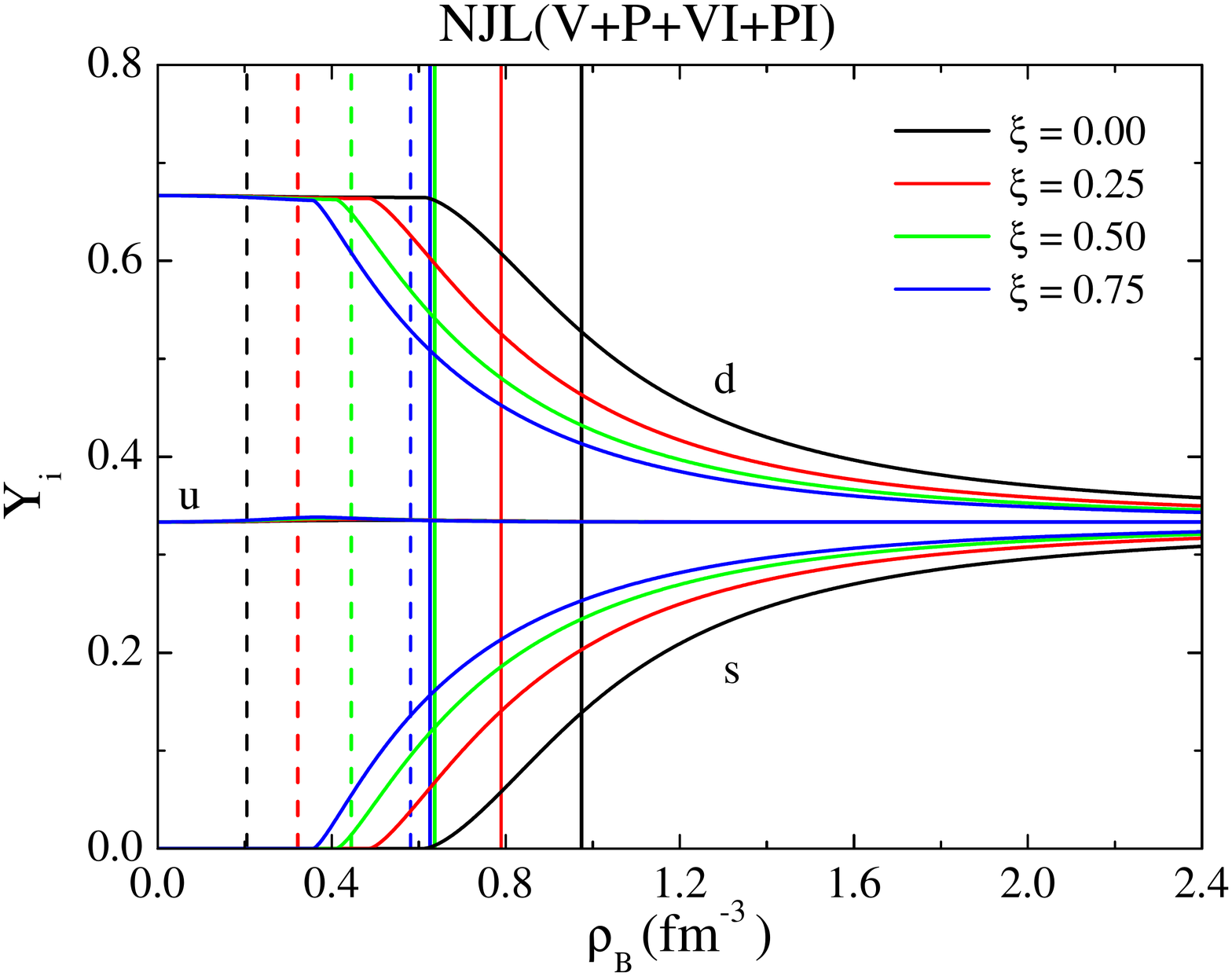}\put(-30,30){\textbf{(a)}}
\hspace{-0.2cm}
\includegraphics[width=0.40\textwidth]{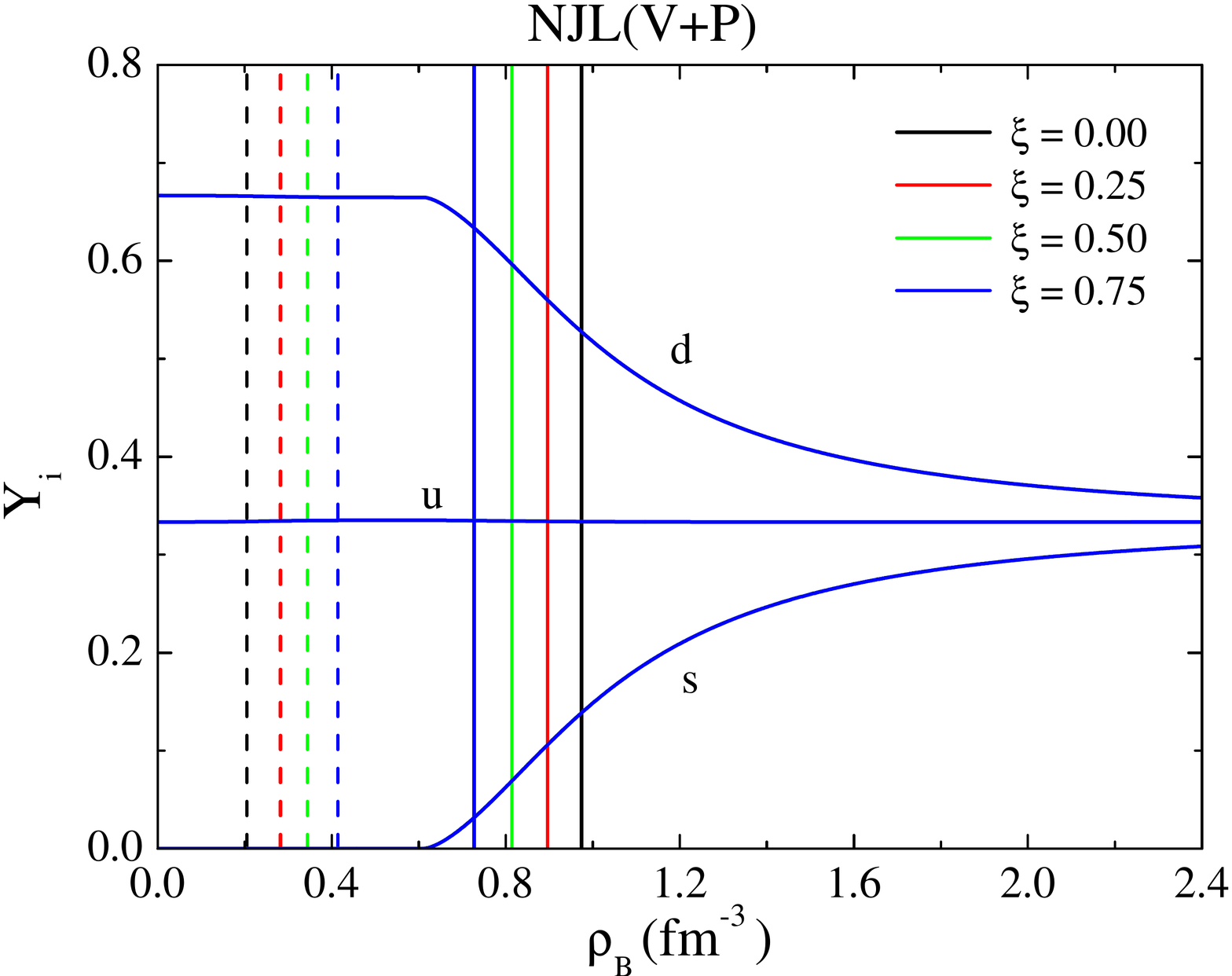}\put(-30,30){\textbf{(b)}}
\hspace{-0.2cm}
\includegraphics[width=0.40\textwidth]{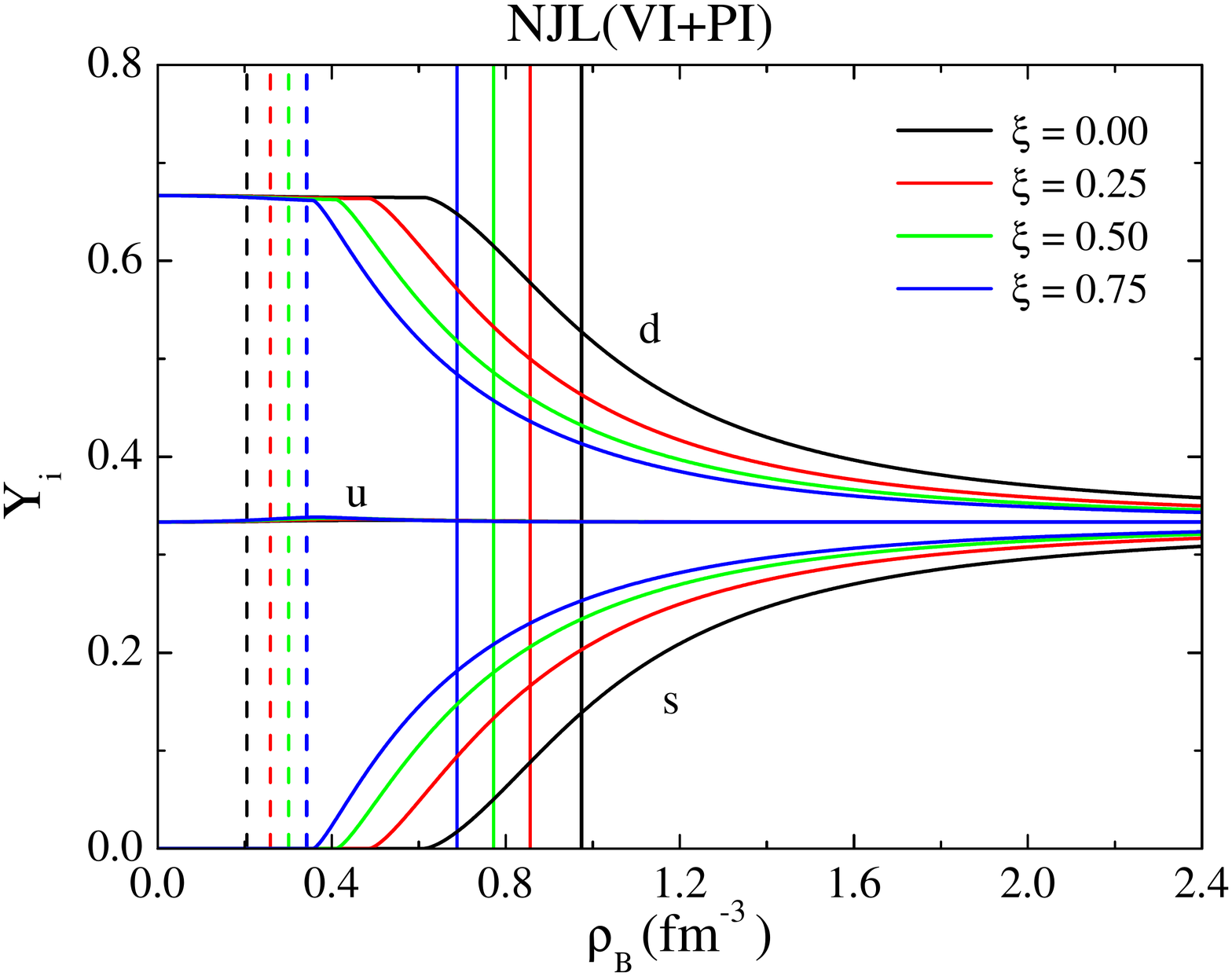}\put(-30,30){\textbf{(c)}}
\caption{Fractions of each flavor of quark ($Y_i$) in function of the baryonic density ($\rho_B$). The central density ($\rho^c$) and initial quark phase density ($\rho^Q$) are shown (full and dashed vertical lines, respectively). The threshold for the emergence of \textit{strange} quarks in the NJL(V+P) model does not depend on $\xi$ ($G_V$) (black line). }
\label{fig:3}
\end{figure}

We will next study the onset of strangeness describing quark matter within the  SU(3) NJL model.
Since the onset of hyperons for NL$3\omega\rho$ occurs at 0.31 fm$^{-3}$ \cite{Fortin:2016hny}, above the onset of quark matter when $B^*$ is included, see Table \ref{tab:8}, except for three cases, we will only consider nucleonic matter in the hadronic phase in order to allow a comparison between parametrizations.
In Fig. \ref{fig:3} the $s$, $d$ and $u$ quark fractions $Y_i=\rho_i/(3\rho_B)$ are plotted. As soon as the $s$-quark sets in the fraction of $d$-quarks suffers a strong reduction, the fractions of $d$ and $s$-quarks approach $\sim 0.33$, asymptotically, the first from above and the second from below. 

Taking the vector-isoscalar interaction alone the strange fraction does not change with $\xi$[see panel (b) in Fig. \ref{fig:3}], which is simply explained because the interaction energy does not depend separately on each flavor \cite{Masuda:2012ed}. 
The vector-isovector interaction distinguishes the flavors and the larger $\xi$ the earlier occurs the $s$-quark onset [see panels (a) and (c) of Fig. \ref{fig:3}]. 
The $u$ quark fraction is practically independent of density, with a value close to 1/3, except for a deviation that can be as high as 0.005 if $\xi=0.75$. 
This deviation from 1/3 is compensated by the presence of electrons in order to turn matter  electrically neutral. 
The onset of strangeness at quite high densities, generally above $3\rho_0\approx0.5 \text{ fm}^{-3}$, is linked to the high constituent mass of the $s$-quark since the partial restoration of chiral symmetry for the $s$-quark occurs at high densities \cite{Buballa:2003qv}.

\begin{table*}[t!]
\begin{ruledtabular}
    \begin{tabular}{cccccccccccc}
    \multirow{2}{*}{Model} & \multirow{2}{*}{$\xi$} & $B^*$ 
    & $\mu_B^{H-Q}$ & $\rho^H$  & $\rho^Q$  & $\rho^c$  & $M_m $  &
    $M_{bm}$  & $R_m $ & $R_{1.4} $ & $N_s/N_B$  \\
    &     & [MeV$\,$fm$^{-3}$]
    & [MeV] & [fm$^{-3}$] & [fm$^{-3}$] & [fm$^{-3}$] & [M$_{\odot}$] & [M$_{\odot}$] & [km] & [km] & [\%] \\
    \hline
    NJL & 0.00  & 0 & 1093 & 0.282 & 0.384 & 0.951 & 1.76  & 2.00  & 11.91 & 13.39  & 1.32 \\ \cline{2-12}
     &  \textbf{0.17}  &  & \textbf{1190} & \textbf{0.338} &  \textbf{0.442} & \textbf{0.734} & \textbf{2.00} & \textbf{2.29} & \textbf{13.08} & \textbf{13.74} & \textbf{0.93}\\
    NJL         & 0.25 & \multirow{2}{*}{0} & 1247   & 0.368 & 0.475 & 0.635 & 2.13  & 2.48  & 13.64  & 13.74  & 0.53 \\
    (V+P+VI+PI) & 0.50 &   & 1410 & 0.444 & 0.640 & 0.578 & 2.47  & 2.94  & 13.96 & 13.74  & 0.04 \\
                & 0.75 &   & 1541 & 0.504 & 0.755 & 0.757 & 2.63  & 3.18  & 13.76 & 13.74  & 0.01 \\ \cline{2-12}
       NJL   &   \textbf{0.25} & \multirow{2}{*}{0} & \textbf{1179}   & \textbf{0.332} & \textbf{0.434} & \textbf{0.816} & \textbf{2.00}  & \textbf{2.30}  & \textbf{12.64}  & \textbf{13.74}  & \textbf{0.50} \\
    (V+P)   & 0.50 &   & 1285 & 0.386 & 0.496 & 0.663 & 2.25  & 2.63  & 13.46 & 13.74  & 0.02 \\
            & 0.75 &   & 1412 & 0.444 & 0.568 & 0.612 & 2.48  & 2.96  & 13.85 & 13.74  & $\sim0$ \\ \cline{2-12}
      NJL   & 0.25  &  & 1147 & 0.314 & 0.416 & 0.766 & 1.83  & 2.08  & 12.88  & 13.74  & 1.80 \\
      (VI+PI)  & 0.50 & \multirow{2}{*}{0} & 1208 & 0.348 & 0.469 & 0.578 & 1.96  & 2.24  & 13.82  & 13.74  & 0.85 \\
     & {\bf 0.60}  &  & {\bf 1225} & {\bf 0.356} & {\bf 0.507} & {\bf 0.429} & {\bf 2.00} & {\bf 2.30} & {\bf 14.00} & {\bf 13.74} & {\bf 0.31}\\
     & 0.75 &   & 1243 & 0.366 & 0.558 & 0.515 & 2.07  & 2.39  & 14.07 &  13.74 & 0.01 \\
    \hline    
    NJL & 0.00  & 6.60 & 999 & 0.198 & 0.205 & 0.974 & 1.78  & 2.05  & 11.55 & 12.33  & 1.43 \\ \cline{2-12}
    & {\bf 0.22}  & {\bf 9.49} & {\bf 1087} & {\bf 0.278} & {\bf 0.315} & {\bf 0.806} & {\bf 2.00} & {\bf 2.29} & {\bf 12.61} & {\bf 13.63} & {\bf 2.07} \\
    NJL & 0.25 & 10.09 & 1100   & 0.286 & 0.322 & 0.789 & 2.02  & 2.33  & 12.73  & 13.71  & 2.14 \\
    (V+P+VI+PI) & 0.50 & 14.62  & 1287 & 0.386 & 0.445 & 0.637 & 2.29  & 2.69  & 13.67 &  13.74 & 1.85 \\
    & 0.75 & 20.57  & 1431 & 0.454 & 0.581 & 0.626 & 2.51  & 3.00  & 13.88 & 13.74  & 0.46 \\ \cline{2-12}
    & 0.25 & 8.61 & 1049 & 0.250 & 0.282 & 0.896 & 1.98  & 2.28  & 12.08 & 13.26  & 0.98 \\    
    NJL & {\bf 0.28}  & {\bf 8.85} & {\bf 1057} & {\bf 0.256} & {\bf 0.290} & {\bf 0.885} & {\bf 2.00} & {\bf 2.31} &  {\bf 12.14} &  {\bf 13.35} &  {\bf 0.91}\\   
    (V+P)  & 0.50 & 10.92 & 1132 & 0.306 & 0.344 & 0.814 & 2.15  & 2.51  & 12.58 & 13.74 & 0.48 \\
           & 0.75 & 13.63 & 1246 & 0.366 & 0.414 & 0.727 & 2.33  & 2.75 & 13.08 &  13.74  & 0.12 \\  \cline{2-12}
  NJL     & 0.25 & 7.92 & 1029 & 0.232 & 0.259 & 0.856 & 1.80  & 2.05  & 12.21 & 12.95  & 2.83 \\
  (VI+PI) & 0.50 & 9.33 & 1072 & 0.268 & 0.301 & 0.772 & 1.81  & 2.06  & 12.75 & 13.52  & 4.12 \\
          & 0.75 & 10.90 & 1129 & 0.304 & 0.342 & 0.688 & 1.84  & 2.08  & 13.24 & 13.74  & 4.77 \\
     \cline{2-12} 
 \Big\{
\begin{tabular}{c}
 $\xi_{\rho}= 0.75$  \\ 
 $\xi_{\omega}=0.15$      
\end{tabular}
&  $-$ & {\bf 12.44} & {\bf 1190} & {\bf 0.338} & {\bf 0.389} & {\bf 0.649} & {\bf 2.00}  & {\bf 2.29} & {\bf 13.52} & {\bf 13.74} & {\bf 3.90}\\ 

    \end{tabular}
\end{ruledtabular}    
    \caption{Baryonic chemical potential ($\mu_B^{H-Q}$), hadron  ($\rho^H$) and quark ($\rho^Q$) baryonic density at  deconfinement and respective value of the parameter $B^*$. Values of central baryonic density ($\rho^c$), maximum gravitational mass ($M_m $), maximum baryonic mass ($M_{bm}$), radius ($R_m $), radius of $1.4M_\odot$ stars ($R_{1.4} $), and the ratio of total number of strange quarks to the total baryon number ($N_s/N_B$) \cite{Hanauske:2001nc} of the respective neutron star, for each model and value of $\xi$, for the SU(3) parameter set.
In bold we present the approximate values of $\xi$ at which $2M_\odot$ are obtained.
The last line corresponds to the combination of $G_{\rho}$ and $G_{\omega}$, in terms of $\xi_{\rho}=G_{\rho}/G_S$ and $\xi_{\omega}=G_{\omega}/G_S$, at which two solar mass are attained.
}
\label{tab:8}
\end{table*}

Properties of hybrid stars, including maximum mass configurations,  
obtained with the SU(3) parametrization are presented in Table 
\ref{tab:8} with $B^*=0$ and  $B^*\ne0$. All  $B^*\ne0$  cases considered show a pure quark matter in the center of the star.
Besides the quantities included in Table \ref{tab:6}, the fraction of strangeness inside the star is also given. If a large $\xi$ parameter is considered the amount of strangeness 
in the star is residual except for the NJL(VI+PI) model: in this case the 
strangeness fraction increases with larger values of $\xi$.

Looking into the details of the NJL(VI+PI) model,  we conclude that when $\xi$ is increased the EoS becomes harder before the onset of strangeness:
the slope of the curve $P$ versus $\rho$ is larger immediately after the hadron-quark transition [see Fig. \ref{fig:2}, panel (c)] allowing stars with a greater mass.
However, the higher $\xi$ the lower the density for the onset of strangeness [as already seen in panel (c) of Fig. \ref{fig:3}].
After the onset of the $s$-quarks, the EoS becomes softer since the Fermi pressure is distributed among a larger number of degrees of freedom.
These two combined effects result in stars with larger masses and  lower central densities, but larger fractions of strangeness.

\begin{figure*}[t!]
\begin{tabular}{cc}
\includegraphics[width=0.43\textwidth]{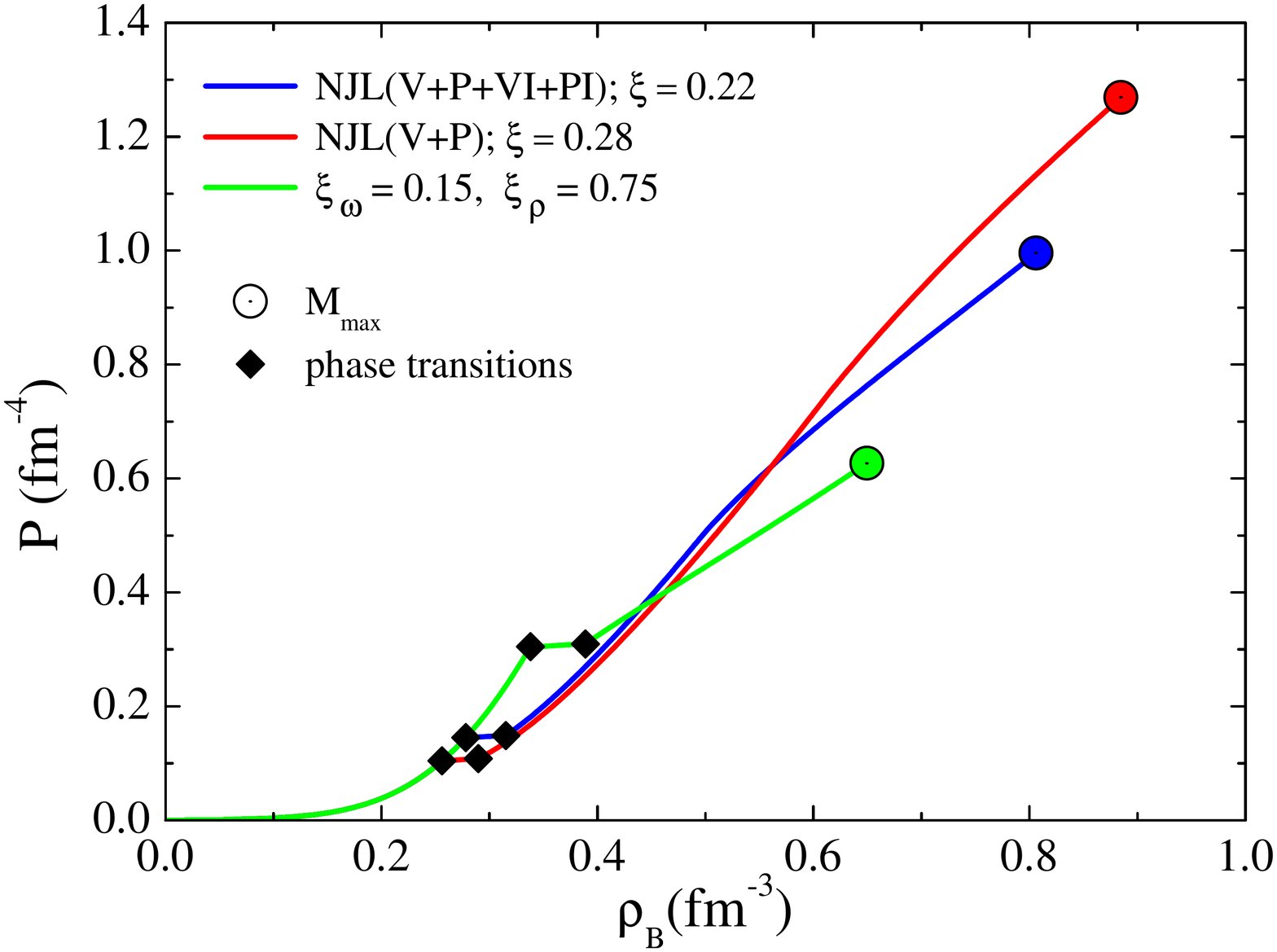}\put(-30,30){\textbf{(a)}}&
\includegraphics[width=0.43\textwidth]{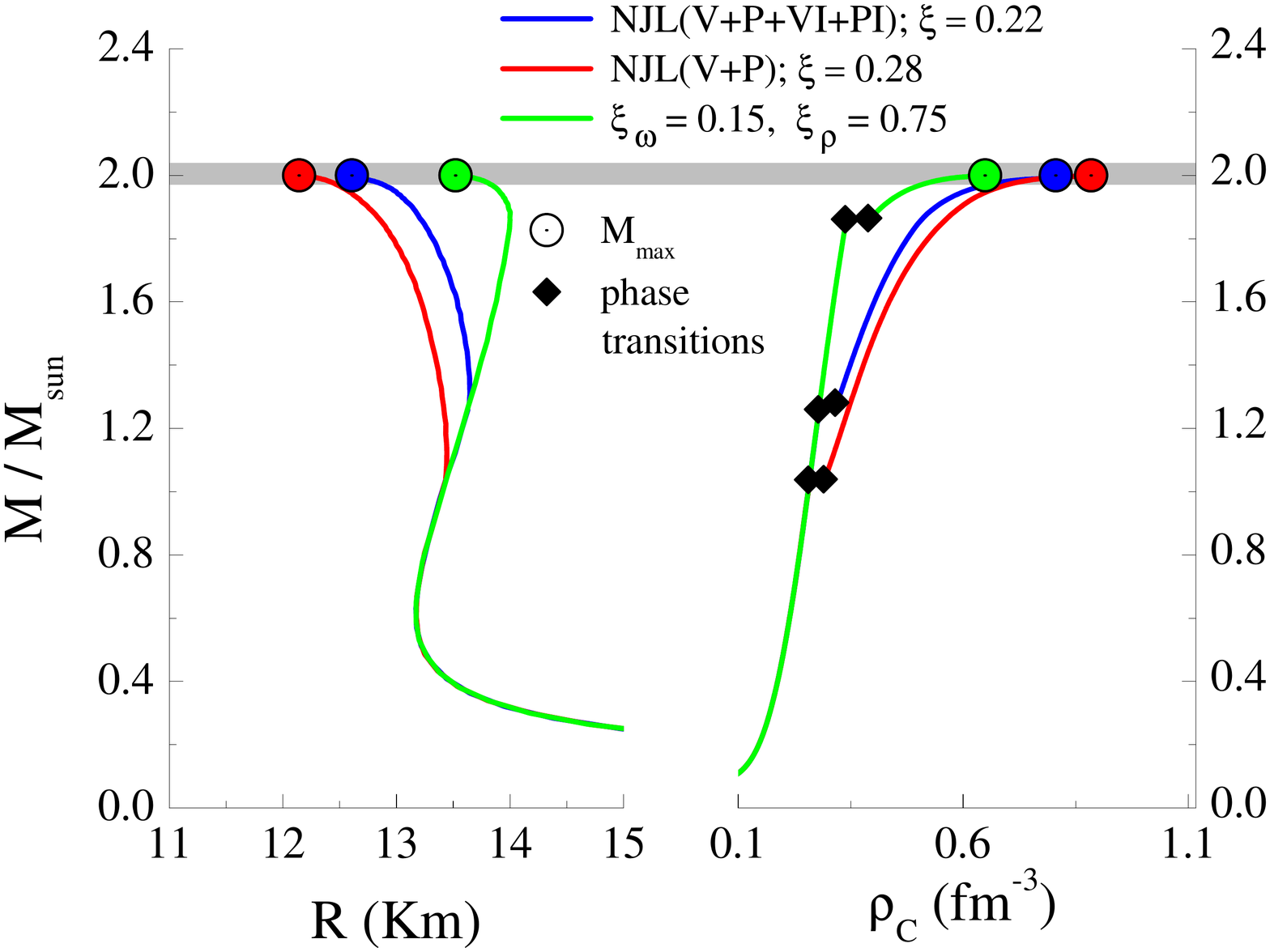}\put(-35,35){\textbf{(b)}}
\end{tabular}
\caption{The EoS (left panel), and the respective  mass-radius curves  (right panel) of the families of stars having a 2$M_\odot$  maximum mass, for three different combinations of $\xi_\omega$ and $\xi_\rho$:  ($\xi_\omega$,$\xi_\rho$)=  (0.28,0) or NJL(V+P) with $\xi=0.28$,  (0.22,0.22) or NJL(V+P+VI+PI) with $\xi=0.22$, and (0.15,0.75).
}
\label{fig:4}
\end{figure*}

Analyzing the radius of the $1.4M_\odot$ stars obtained within the different parametrizations, see Tables \ref{tab:6} and \ref{tab:8}, we conclude that  most of these stars have $R=13.74$ km corresponding to hadronic stars with no quark content. 
However, some models with  $B^*\ne 0$ predict the existence of quark matter  inside low mass stars with $M<1.4M_\odot$. These stars  have the particularity of having smaller radii. 
In fact, it is possible to get $1.4M_\odot$ stars with $R<13.74$ km within families that predict $2M_\odot$ stars. For SU(3) NJL, the smallest radius obtained is $13.35$ km above the $10.1 - 11.1$ km prediction of \cite{Ozel:2015fia} from the analysis of spectroscopic radius measurements during thermonuclear bursts or in quiescence or even the 12.1$\pm$1.1 km obtained in \cite{Steiner:2015aea} from experimental constraints and causality restrictions. However, in  \cite{suleimanov16} radii above 13 km were obtained for X-ray bursting NS and  in \cite{Chen:2015zpa} it has been shown that causality together with the $2M_\odot$ constraints imposes $R>10.7$ km. For a recent review of  the current status of measurements of radius of neutron stars see \cite{haensel16}.
Stronger constraints on neutron star radii are expected from future X-ray telescopes like NICER and Athena.
The measurement of the radius of low mass stars such as the pulsar  PSR  J1918-0642 with a mass the $1.18 ^{+0.10}_{-0.11} M_\odot$ could give some indication  on the properties of the EOS at densities just above saturation density and constrain the onset density of quark matter. In the present calculation it is seen that an early onset gives rise to smaller low-mass star radii. However, the radii differences with respect to pure nucleonic matter are probably not strong enough to allow conclusive results mainly because the hadronic EOS itself  has still large uncertainties at those densities.

Finally, we present the results for the combination of $G_{\rho}$ and $G_{\omega}$, in terms of $\xi_{\rho}=G_{\rho}/G_S$ and $\xi_{\omega}=G_{\omega}/G_S$, for which $2.0\,M_\odot$ are obtained: $\xi_{\rho}=0.75$ and $\xi_{\omega}=0.15\,\,$\footnote{By fixing $\xi_{\rho}=0.75$ with $\xi_{\omega}=0$ we have the model NJL (VI+PI) for $\xi=0.75$.}. This will allow us to clarify some aspects reported previously.

When the vector-isovector interaction is absent ($\xi_{\rho}=0$), the EoS is harder at high densities (see red curve in Fig. \ref{fig:4}, left panel) because the fraction of strangeness is very low. 
When $\xi_{\rho}$ and $\xi_{\omega}$ are mixed, the larger  $\xi_{\rho}$, the lower the onset density of strangeness and, therefore, the larger the fraction of strange quarks. Simultaneously the hadron-quark transition occurs at higher densities and the central densities decreases: the larger $s$-quark contribution softens the quark EoS, and, in order to attain the 2$M_\odot$ the contribution of the hadronic star component has to be larger.
For example, taking $\xi_{\rho}=0.75$ and $\xi_{\omega}=0.15$, $\mu_B^{H-Q}$ has the highest value when compared with NJL(V+P) model for $\xi=0.28$ and with NJL(V+P+VI+PI) model for $\xi=0.22$, while $\rho^c$ has the smallest value, as it can be seen in Table \ref{tab:8} and in Fig. \ref{fig:4}, right panel (for all three cases the maximum gravitational mass is 2$M_{\odot}$).

Due to the lack of strangeness in the SU(2) case, the influence of vector-isovector interaction is much smaller when compared with vector-isoscalar interaction. Taking the 2$M_{\odot}$ cases in Table \ref{tab:6} it can be seen that the hadron-quark phase transition, and the star properties, are very close for cases with vector-isoscalar interaction [NJL(V+P+VI+PI) and NJL(VI+PI) models]. To have a 2$M_{\odot}$ star with a vector-isovector it is needed a much stronger coupling, however, the hadron-quark phase transition and the star properties are not very different from the other cases (see Table \ref{tab:6}). 


\section{Conclusions}
\label{conclusions}

We have analyzed the possibility of obtaining hybrid stars with the quark core described within the NJL model with and without strangeness content. 
Earlier works have shown that only under some conditions a pure quark matter core occurs when quark matter is described within this  model.
It is, therefore, important to  choose adequately the properties of the hadron and the quark phases. In the present work, besides considering the coincidence between the deconfinement phase transition and the partial restoration of chiral symmetry, two new parametrizations of the SU(2) and SU(3) NJL models are proposed with a low vacuum constituent quark mass equal to 313 MeV. 
As shown in \cite{Buballa:2003et} a smaller vacuum constituent quark mass favors a hadron-quark phase transition at lower densities and stable stars with a quark core.

We have considered together with the usual scalar and pseudoscalar terms in the NJL model also vector-isoscalar and vector-isovector terms. The  vector-isoscalar terms  have an important effect on the order of the chiral phase transition and turn the EoS harder \cite {Hanauske:2001nc,Pagliara:2007ph,Bonanno:2011ch}.
This, in fact, is also true for the vector-isovector terms,  although the EoS does not become so hard and smaller maximum mass configurations are obtained. The inclusion of a vector-isovector term allows larger quark cores, the onset of quark matter at lower densities, smaller hadron-quark mixed phases,  and, in the SU(3) version, a larger strangeness content for the same coupling strength.
A larger vector-isovector  coupling shifts the deconfinement to larger densities and gives rise to a smaller quark contribution to the hybrid star properties, mainly if the vector-isoscalar is also considered. 

We studied the possibility of getting 2$M_\odot$ stars including both vector-isoscalar and vector-isovector terms.
It was shown that for  the SU(3) NJL  2$M_\odot$ configurations always require the presence of a  vector-isoscalar term, and that the larger the  vector-isovector term the larger the strangeness fraction  but the larger the hadron-quark transition density and, therefore, the  smaller the quark contribution to the star. It is the $s$-quark with its quite high mass that causes this behavior.  In the case of SU(2) NJL, properties of the 2$M_\odot$ stars taking different strengths for the vector-isoscalar and isovector terms are almost indistinguishable.

In the present work we have fixed the bag term $B^*$ imposing that the deconfinement and the chiral phase transitions coincide. Presently, it is still not clear if both phase transitions coincide, and other scenarios are possible, such as a chiral symmetry restoration before the deconfinement is attained, giving rise to a quarkyonic phase. Imposing different constraints on the $B^*$ will have essentially quantitative effects, shifting the onset of quark matter and giving rise to a smaller or larger density jump at the first-order phase transition, but the qualitative features are similar to the ones discussed imposing the coincidence of the chiral and deconfinement transitions.  

The main conclusion of the present work is the importance of choosing conveniently the quark model parameters when building a hadron-quark EoS. We have shown that fixing the vacuum quark constituent mass with a value that is one third of the vacuum nucleon mass and, therefore, a baryonic chemical potential at zero density in the quark phase  equal to the one in the hadronic phases allows the appearance of a pure quark core in the center of a neutron star. Including a strong enough  vector-isoscalar interaction will result in maximum mass configurations with masses above $2M_\odot$. With a vector-isovector interaction alone this is not  possible within the SU(3) NJL model, on the other hand, this interaction causes a larger strangeness content and a softening of the quark EoS.
However, as in previous studies that have included the strangeness degree of freedom, the strangeness content of these stars  is generally very small.

\vspace{1cm}

{\bf Acknowledgment: } 
This work was supported by NewCompStar, COST Action MP1304, and by FCT (Fundação para a Ciência e Tecnologia), Portugal, under the Grant No. SFRH/BPD/102273/2014 (P. C.), and under the project No. UID/FIS/04564/2016.
\vspace{0.5cm}


\appendix
\section{Quark phase equation of state}
\label{Appendix} 

\subsection{Quark chemical potentials in SU(2) and SU(3)}

The expressions for the chemical potentials in SU(3), defined in Eq. (\ref{L_vect}), are given by:
\begin{itemize}
\item[--] for NJL(V+P+VI+PI), when $G_\omega=G_\rho=G_V$,
\begin{equation}
\tilde{\mu}_i=\mu_i-4G_V\rho_i;\,\,\, i=u,d,s;
\label{su3V+VI}
\end{equation}
\item[--] for NJL(V+P), when $G_\rho=0$ and $G_\omega=G_V$,
\begin{align}
\tilde{\mu}_i&=\mu_i -\frac{4}{3}G_V \left( \rho_i + \rho_j + \rho_k\right),\\
&i\ne j\ne k\in \{u,d,s\};\nonumber
\label{su3V}
\end{align}
\item[--] for NJL(VI+PI), when $G_\omega=0$ and $G_\rho=G_V$,
\begin{align}
\tilde{\mu}_i&=\mu_i-\frac{4}{3}G_V\left( 2\rho_i - \rho_j - \rho_k\right),\\
&i\ne j\ne k\in \{u,d,s\}.\nonumber
\label{su3VI}
\end{align}
\end{itemize}

The expressions for the chemical potentials in SU(2), defined in Eq. (\ref{L_vect}), are given by:
\begin{itemize}
\item[--] for NJL(V+P+VI+PI), when $G_\omega=G_\rho=G_V$,
\begin{equation}
\tilde{\mu}_i=\mu_i-4G_V\rho_i,\;\;i \in \{ u,d\};
\label{su2V+VI}
\end{equation}
\item[--] for NJL(V+P), when $G_\rho=0$ and $G_\omega=G_V$,
\begin{equation}
\tilde{\mu}_i=\mu_i-2G_V\left( \rho_i + \rho_j \right),\;\;i\neq j  \in \{ u,d\};
\label{su2V}
\end{equation}
\item[--] for NJL(VI+PI), when $G_\omega=0$ and $G_\rho=G_V$,
\begin{equation}
\tilde{\mu}_i=\mu_i-2G_V\left( \rho_i - \rho_j \right),\;\;i\neq j  \in \{ u,d\}.
\label{su2VI}
\end{equation}
\end{itemize}


\subsection{Thermodynamic quantities in SU(2) and SU(3)}
In SU$(2)$ as well as in SU$(3)$, the quark condensate for each flavor is given by:
\begin{equation}
\sigma_i=\braket{\bar{q}_i q_i}=-2N_c\int \frac{d^3p}{(2\pi)^3}\frac{M_i}{E_i} \left( 1-n_i-\bar{n}_i \right)
\end{equation}
where $n_i$ and $\bar{n}_i$ are the quark and anti-quark occupation numbers:
\begin{align}
n_i&=\frac{1}{e^{(E_i-\tilde{\mu}_i)/T}+1} \\  
\bar{n}_i&=\frac{1}{e^{(E_i+\tilde{\mu}_i)/T}+1}
\end{align}
The $i-$quark number density, $\rho_i = - (\partial\Omega/\partial\mu_i)$ reads
\begin{equation}
\rho_i=2N_c\int \frac{d^3p}{(2\pi)^3}\left(n_i-\bar{n}_i \right).
\end{equation}
The leptonic contribution ($\beta$-Equilibrium) to the pressure is
\begin{widetext}
\begin{equation}
P^{\,\beta\text{-eq.}} = P_{\text{ NJL}} + 2T \int \frac{d^3p}{(2\pi)^3}  \Big[  \ln \big( 1 + e^{-(E_e+\mu_e)/T}  \big) + \ln \big( 1 + e^{-(E_e-\mu_e)/T}  \big) \Big],
\end{equation}
\end{widetext}
being $E_e=\sqrt{p^2+m_e^2}$, and to the energy density is
\begin{equation}
\epsilon^{\,\beta\text{-eq.}} = \epsilon_{\text{NJL}} -2\int \frac{d^3p}{(2\pi)^3}  E_e(n_e+\bar{n}_e),
\end{equation}
where $n_e$ and $\bar{n}_e$ are, respectively,
\begin{align}
n_e&=\frac{1}{e^{(E_e-\mu_e)/T}+1}   \\
\bar{n}_e&=\frac{1}{e^{(E_e+\mu_e)/T}+1}.
\end{align}
The electron density ($\rho_e = - (\partial\Omega_e/\partial\mu_e)$) is given by 
\begin{equation}
\rho_e=2\int \frac{d^3p}{(2\pi)^3}\left(n_e-\bar{n}_e \right),
\end{equation}

In the limit $T=0$:
\begin{equation}
\sigma_i=\braket{\bar{q}_i q_i}=-\frac{N_c}{\pi^2}\int^{\Lambda}_{\lambda_{F_i}} \text{d}p p^2 \frac{M_i}{\sqrt{p^2+M_i^2}},
\end{equation}
with the Fermi momentum of the respective quark flavor $i$ given by 
\begin{equation}
\lambda_{F_i}=\sqrt{\tilde{\mu_i}^2 - M_i^2 },
\end{equation}
and the density given by 
\begin{equation}
\rho_i=\frac{N_c}{\pi^2} \frac{\lambda_{F_i}^3}{3}.
\label{densT0}
\end{equation}
For electrons it comes:
\begin{equation}
\lambda_{F_e}=\sqrt{\tilde{\mu_e}^2 - m_e^2 },
\end{equation}
and
\begin{equation}
\rho_e= \frac{\lambda_{F_i}^3}{3\pi^2}.
\end{equation}

\subsubsection{SU(2)}

\begin{widetext}

The pressure and energy density in SU(2) are respectively given by:
\begin{align}
P_{\text{NJL}}  & = - \Omega_0 -  G_S \left( \sigma_u + \sigma_d \right)^2 + G_\omega \left( \rho_u + \rho_d \right)^2 + G_\rho \left( \rho_u - \rho_d \right)^2  \nonumber
\\
& + 2 N_c \sum_{i=u,d} \int \frac{d^3p}{(2\pi)^3} \left[  E_i + T \ln \left( 1 + e^{-(E_i+\tilde{\mu}_i)/T}  \right) + T \ln \left( 1 + e^{-(E_i-\tilde{\mu}_i)/T}  \right) \right] ,
\end{align}
and
\begin{align}
\epsilon_{\text{NJL}} & = \Omega_0 +  G_S \left( \sigma_u + \sigma_d \right)^2 - G_\omega \left( \rho_u + \rho_d \right)^2 - G_\rho \left( \rho_u - \rho_d \right)^2 \nonumber
\\
& - 2 N_c \sum_{i=u,d} \int \frac{d^3p}{(2\pi)^3} \left[ E_i \left( 1 - n_i - \bar{n}_i \right) + n_i\left(\tilde{\mu}_i - \mu_i\right) + \bar{n}_i\left( \mu_i - \tilde{\mu}_i \right) \right] .   
\end{align}

In the limit $T=0$ the pressure is given by
\begin{align}
P_{\text{NJL}}   &= - \Omega_0  -  G_S \left( \sigma_u + \sigma_d \right)^2 + G_\omega \left( \rho_u + \rho_d \right)^2 + G_\rho \left( \rho_u - \rho_d \right)^2  + \frac{N_c}{\pi^2}\sum_{i=u,d} \int^{\Lambda}_{\lambda_{F_i}} dp  p^2 E_i  + \frac{N_c}{\pi^2}\sum_{i=u,d} \tilde{\mu}_i \frac{\lambda_{F_i}^3}{3}  ,
\end{align}
where the quark density of flavor $f$ is given by Eq. (\ref{densT0}),
and energy density is given by
\begin{align}
\epsilon_{\text{NJL}} &=  \Omega_0 +  G_S \left( \sigma_u + \sigma_d \right)^2 - G_\omega \left( \rho_u + \rho_d \right)^2 - G_\rho \left( \rho_u - \rho_d \right)^2 -\frac{N_c}{\pi^2}\sum_{i=u,d} \int^{\Lambda}_{\lambda_{F_i}} dp  p^2 E_i  + \frac{N_c}{\pi^2}\sum_{i=u,d} \left(\mu_i -\tilde{\mu}_i \right) \frac{\lambda_{F_i}^3}{3}
\end{align}

\subsubsection{SU(3)}

The pressure and energy density in SU$(3)$ are given by:
\begin{align}
P_{\text{NJL}} & = - \Omega_0 - 2 G_S \left( \sigma_u^2 + \sigma_d^2 + \sigma_s^2 \right) + 4 G_D \sigma_u \sigma_d \sigma_s    \nonumber
\\
& + \frac{2}{3} G_\omega \left( \rho_u + \rho_d + \rho_s \right)^2 + G_\rho \left( \rho_u - \rho_d \right)^2 
+ \frac{1}{3} G_\rho \left(  \rho_u + \rho_d - 2 \rho_s \right)^2 \nonumber
\\
& + 2 N_c \sum_{i=u,d,s} \int \frac{d^3p}{(2\pi)^3} \left[  E_i + T \ln \left( 1 + e^{-(E_i+\tilde{\mu}_i)/T}  \right) + T \ln \left( 1 + e^{-(E_i-\tilde{\mu}_i)/T}  \right) \right] , 
\end{align}
and
\begin{align}
\epsilon_{\text{NJL}} & = \Omega_0 + 2 G_S \left( \sigma_u^2 + \sigma_d^2 + \sigma_s^2 \right) - 4 G_D \sigma_u \sigma_d \sigma_s    \nonumber
\\
& - \frac{2}{3} G_\omega \left( \rho_u + \rho_d + \rho_s \right)^2 - G_\rho \left( \rho_u - \rho_d \right)^2 
- \frac{1}{3} G_\rho \left(  \rho_u + \rho_d - 2 \rho_s \right)^2 \nonumber
\\
& -2 N_c \sum_{i=u,d,s} \int \frac{d^3p}{(2\pi)^3} \left[ E_i \left( 1 - n_i - \bar{n}_i \right) + n_i\left(\tilde{\mu}_i - \mu_i\right) + \bar{n}_i\left( \mu_i - \tilde{\mu}_i \right) \right] . 
\end{align}

In the limit $T=0$ the pressure becomes,
\begin{align}
P_{\text{NJL}}   &= - \Omega_0 - 2 G_S \left( \sigma_u^2 + \sigma_d^2 + \sigma_s^2 \right) + 4 G_D \sigma_u \sigma_d \sigma_s  
+ \frac{N_c}{\pi^2}\sum_{i=u,d,s} \int^{\Lambda}_{\lambda_{F_i}} dp  p^2 E_i  + \frac{N_c}{\pi^2}\sum_{i=u,d} \tilde{\mu}_i \frac{\lambda_{F_i}^3}{3}  \nonumber  
\\
& + \frac{2}{3} G_\omega \left( \rho_u + \rho_d + \rho_s \right)^2 + G_\rho \left( \rho_u - \rho_d \right)^2 
+ \frac{1}{3} G_\rho \left(  \rho_u + \rho_d - 2 \rho_s \right)^2,
\end{align}
and energy density is,
\begin{align}
\epsilon_{\text{NJL}} &=  \Omega_0 + 2 G_S \left( \sigma_u^2 + \sigma_d^2 + \sigma_s^2 \right) - 4 G_D \sigma_u \sigma_d \sigma_s    
 - \frac{N_c}{\pi^2}\sum_{i=u,d,s} \int^{\Lambda}_{\lambda_{F_i}} dp  p^2 E_i  + \frac{N_c}{\pi^2}\sum_{i=u,d} \left(\mu_i -\tilde{\mu}_i \right) \frac{\lambda_{F_i}^3}{3} \nonumber
\\
&  - \frac{2}{3} G_\omega \left( \rho_u + \rho_d + \rho_s \right)^2 - G_\rho \left( \rho_u - \rho_d \right)^2 
- \frac{1}{3} G_\rho \left(  \rho_u + \rho_d - 2 \rho_s \right)^2.
\end{align}
\end{widetext}


\begin{thebibliography}{99}

\bibitem{Antoniadis:2013pzd} 
  J.~Antoniadis {\it et al.},
  Science {\bf 340}, 6131 (2013),
  doi:10.1126/science.1233232,
  [arXiv:1304.6875 [astro-ph.HE]].



\bibitem{Demorest:2010bx} 
  P.~Demorest, T.~Pennucci, S.~Ransom, M.~Roberts and J.~Hessels,
  Nature {\bf 467}, 1081 (2010),
  doi:10.1038/nature09466,
  [arXiv:1010.5788 [astro-ph.HE]].



\bibitem{Fonseca:2016tux} 
  E.~Fonseca {\it et al.},
  arXiv:1603.00545 [astro-ph.HE].



\bibitem{Glendenning}
	N. K. Glendenning, 
  {\it Compact Stars} (Springer-Verlag, New York, 2000).

\bibitem{Chodos:1974pn} 
  A.~Chodos, R.~L.~Jaffe, K.~Johnson and C.~B.~Thorn,
  Phys.\ Rev.\ D {\bf 10}, 2599 (1974),
  doi:10.1103/PhysRevD.10.2599.



\bibitem{Bombaci:2004mt} 
  I.~Bombaci, I.~Parenti and I.~Vidana,
  Astrophys.\ J.\  {\bf 614}, 314 (2004),
  doi:10.1086/423658,
  [astro-ph/0402404].



\bibitem{Schertler:1999xn} 
  K.~Schertler, S.~Leupold and J.~Schaffner-Bielich,
  Phys.\ Rev.\ C {\bf 60}, 025801 (1999),
  doi:10.1103/PhysRevC.60.025801,
  [astro-ph/9901152].



\bibitem{Baldo:2002ju} 
  M.~Baldo, M.~Buballa, F.~Burgio, F.~Neumann, M.~Oertel and H.~J.~Schulze,
  Phys.\ Lett.\ B {\bf 562}, 153 (2003),
  doi:10.1016/S0370-2693(03)00556-2,
  [nucl-th/0212096].



\bibitem{Menezes:2003xa} 
  D.~P.~Menezes and C.~Providencia,
  Phys.\ Rev.\ C {\bf 68}, 035804 (2003),
  doi:10.1103/PhysRevC.68.035804,
  [nucl-th/0308041].



\bibitem{Shovkovy:2003ce} 
  I.~Shovkovy, M.~Hanauske and M.~Huang,
  Phys.\ Rev.\ D {\bf 67}, 103004 (2003),
  doi:10.1103/PhysRevD.67.103004,
  [hep-ph/0303027].



\bibitem{Buballa:2003et} 
  M.~Buballa, F.~Neumann, M.~Oertel and I.~Shovkovy,
  Phys.\ Lett.\ B {\bf 595}, 36 (2004),
  doi:10.1016/j.physletb.2004.05.064,
  [nucl-th/0312078].


\bibitem{Pagliara:2007ph} 
  G.~Pagliara and J.~Schaffner-Bielich,
  Phys.\ Rev.\ D {\bf 77}, 063004 (2008),
  doi:10.1103/PhysRevD.77.063004,
  [arXiv:0711.1119 [astro-ph]].



\bibitem{Buballa:1998pr} 
  M.~Buballa and M.~Oertel,
  Phys.\ Lett.\ B {\bf 457}, 261 (1999),
  doi:10.1016/S0370-2693(99)00533-X,
  [hep-ph/9810529].



\bibitem{Bonanno:2011ch} 
  L.~Bonanno and A.~Sedrakian,
  Astron.\ Astrophys.\  {\bf 539}, A16 (2012),
  doi:10.1051/0004-6361/201117832,
  [arXiv:1108.0559 [astro-ph.SR]].



\bibitem{Logoteta:2013ipa} 
  D.~Logoteta, C.~Providência and I.~Vidaña,
  Phys.\ Rev.\ C {\bf 88}, no. 5, 055802 (2013),
  doi:10.1103/PhysRevC.88.055802,
  [arXiv:1311.0618 [nucl-th]].


\bibitem{EPJA2016}
 Blaschke, David, Schaffner-Bielich, Jürgen, and Schulze,
Hans-Josef, Eur. Phys. J. A {\bf 52}, 71 (2016), 
doi:10.1140/epja/i2016-16071-8.

\bibitem{Burgio:2015zka} 
  G.~F.~Burgio and D.~Zappalà,
  Eur.\ Phys.\ J.\ A {\bf 52}, no. 3, 60 (2016),
  doi:10.1140/epja/i2016-16060-y,
  [arXiv:1509.00841 [nucl-th]].

\bibitem{Fraga:2015xha} 
  E.~S.~Fraga, A.~Kurkela and A.~Vuorinen,
  Eur.\ Phys.\ J.\ A {\bf 52}, no. 3, 49 (2016),
  doi:10.1140/epja/i2016-16049-6,
  [arXiv:1508.05019 [nucl-th]].

\bibitem{Drago:1995pr} 
  A.~Drago, M.~Fiolhais and U.~Tambini,
  Nucl.\ Phys.\ {\bf A588}, 801 (1995),
  doi:10.1016/0375-9474(95)00076-D,
  [hep-ph/9503462].

\bibitem{Logoteta:2012zz} 
  D.~Logoteta, I.~Bombaci, C.~Providencia and I.~Vidana,
  Phys.\ Rev.\ D {\bf 85}, 023003 (2012),
  doi:10.1103/PhysRevD.85.023003,
  [arXiv:1203.4159 [nucl-th]].

\bibitem{Alvarez-Castillo:2016oln} 
  D.~Alvarez-Castillo, A.~Ayriyan, S.~Benic, D.~Blaschke, H.~Grigorian and S.~Typel,
  Eur.\ Phys.\ J.\ A {\bf 52}, no. 3, 69 (2016),
  doi:10.1140/epja/i2016-16069-2,
  [arXiv:1603.03457 [nucl-th]].

\bibitem{Lawley:2006ps} 
  S.~Lawley, W.~Bentz and A.~W.~Thomas,
  J.\ Phys.\ G {\bf 32}, 667 (2006),
  doi:10.1088/0954-3899/32/5/006,
  [nucl-th/0602014].

\bibitem{Pais:2016dng} 
  H.~Pais, D.~P.~Menezes and C.~Providência,
  Phys.\ Rev.\ C {\bf 93}, no. 6, 065805 (2016),
  doi:10.1103/PhysRevC.93.065805,
  [arXiv:1603.01239 [nucl-th]].

\bibitem{Hanauske:2001nc} 
  M.~Hanauske, L.~M.~Satarov, I.~N.~Mishustin, H.~Stoecker and W.~Greiner,
  Phys.\ Rev.\ D {\bf 64}, 043005 (2001),
  doi:10.1103/PhysRevD.64.043005,
  [astro-ph/0101267].

\bibitem{Klahn:2006iw} 
  T.~Klahn, D.~Blaschke, F.~Sandin, C.~Fuchs, A.~Faessler, H.~Grigorian, G.~Ropke and J.~Trumper,
  Phys.\ Lett.\ B {\bf 654}, 170 (2007),
  doi:10.1016/j.physletb.2007.08.048,
  [nucl-th/0609067].

\bibitem{Lenzi:2012xz} 
  C.~H.~Lenzi and G.~Lugones,
  Astrophys.\ J.\  {\bf 759}, 57 (2012),
  doi:10.1088/0004-637X/759/1/57,
  [arXiv:1206.4108 [astro-ph.SR]].

\bibitem{Masuda:2012ed} 
  K.~Masuda, T.~Hatsuda and T.~Takatsuka,
  PTEP {\bf 2013}, no. 7, 073D01 (2013),
  doi:10.1093/ptep/ptt045,
  [arXiv:1212.6803 [nucl-th]].

\bibitem{Klahn:2013kga} 
  T.~Klähn, R.~Łastowiecki and D.~B.~Blaschke,
  Phys.\ Rev.\ D {\bf 88}, no. 8, 085001 (2013),
  doi:10.1103/PhysRevD.88.085001,
  [arXiv:1307.6996].

\bibitem{Menezes:2014aka} 
  D.~P.~Menezes, M.~B.~Pinto, L.~B.~Castro, P.~Costa and C.~Providência,
  Phys.\ Rev.\ C {\bf 89}, no. 5, 055207 (2014),
  doi:10.1103/PhysRevC.89.055207,
  [arXiv:1403.2502 [nucl-th]].

\bibitem{Klahn:2015mfa} 
  T.~Klahn and T.~Fischer,
  Astrophys.\ J.\  {\bf 810}, no. 2, 134 (2015),
  doi:10.1088/0004-637X/810/2/134,
  [arXiv:1503.07442 [nucl-th]].

\bibitem{Ferrer:2015vca} 
  E.~J.~Ferrer, V.~de la Incera and L.~Paulucci,
  Phys.\ Rev.\ D {\bf 92}, no. 4, 043010 (2015),
  doi:10.1103/PhysRevD.92.043010,
  [arXiv:1501.06597 [hep-ph]].

\bibitem{Fukushima:2008wg} 
  K.~Fukushima,
  Phys.\ Rev.\ D {\bf 77}, 114028 (2008),
  Erratum: [Phys.\ Rev.\ D {\bf 78}, 039902 (2008)],
  doi:10.1103/PhysRevD.77.114028, 10.1103/PhysRevD.78.039902,
  [arXiv:0803.3318 [hep-ph]].

\bibitem{Klimt:1989pm} 
  S.~Klimt, M.~F.~M.~Lutz, U.~Vogl and W.~Weise,
  Nucl.\ Phys.\  {\bf A516}, 429 (1990),
  doi:10.1016/0375-9474(90)90123-4.

\bibitem{Lutz:1992dv} 
  M.~F.~M.~Lutz, S.~Klimt and W.~Weise,
  Nucl.\ Phys.\  {\bf A542}, 521 (1992),
  doi:10.1016/0375-9474(92)90256-J.


\bibitem{Bratovic:2012qs} 
  N.~M.~Bratovic, T.~Hatsuda and W.~Weise,
  Phys.\ Lett.\ B {\bf 719}, 131 (2013),
  doi:10.1016/j.physletb.2013.01.003,
  [arXiv:1204.3788 [hep-ph]].


\bibitem{Lourenco:2012yv} 
  O.~Lourenco, M.~Dutra, T.~Frederico, A.~Delfino and M.~Malheiro,
  Phys.\ Rev.\ D {\bf 85}, 097504 (2012),
  doi:10.1103/PhysRevD.85.097504,
  [arXiv:1204.6357 [nucl-th]].

\bibitem{Rehberg:1995kh} 
  P.~Rehberg, S.~P.~Klevansky and J.~Hufner,
  Phys.\ Rev.\ C {\bf 53}, 410 (1996),
  doi:10.1103/PhysRevC.53.410,
  [hep-ph/9506436].


\bibitem{Horowitz:2000xj} 
  C.~J.~Horowitz and J.~Piekarewicz,
  Phys.\ Rev.\ Lett.\  {\bf 86}, 5647 (2001),
  doi:10.1103/PhysRevLett.86.5647,
  [astro-ph/0010227].


\bibitem{Fortin:2016hny} 
  M.~Fortin, C.~Providencia, A.~R.~Raduta, F.~Gulminelli, J.~L.~Zdunik, P.~Haensel and M.~Bejger,
  arXiv:1604.01944 [astro-ph.SR].


\bibitem{Costa:2005cz} 
  P.~Costa, M.~C.~Ruivo, C.~A.~de Sousa and Y.~L.~Kalinovsky,
  Phys.\ Rev.\ D {\bf 71}, 116002 (2005),
  doi:10.1103/PhysRevD.71.116002,
  [hep-ph/0503258].

\bibitem{Buballa:2003qv} 
  M.~Buballa,
  Phys.\ Rept.\  {\bf 407}, 205 (2005),
  doi:10.1016/j.physrep.2004.11.004,
  [hep-ph/0402234].


\bibitem{Moreira:2010bx} 
  J.~Moreira, B.~Hiller, A.~A.~Osipov and A.~H.~Blin,
  Int.\ J.\ Mod.\ Phys.\ A {\bf 27}, 1250060 (2012),
  doi:10.1142/S0217751X12500601,
  [arXiv:1008.0569 [hep-ph]].

\bibitem{Agashe:2014kda} 
  K.~A.~Olive {\it et al.} [Particle Data Group Collaboration],
  Chin.\ Phys.\ C {\bf 38}, 090001 (2014),
  doi:10.1088/1674-1137/38/9/090001.



\bibitem{Costa:2010zw} 
  P.~Costa, M.~C.~Ruivo, C.~A.~de Sousa and H.~Hansen,
  Symmetry {\bf 2}, 1338 (2010),
  doi:10.3390/sym2031338,
  [arXiv:1007.1380 [hep-ph]].



\bibitem{tov}
	J. R. Oppenheimer and G. M. Volkoff, 
  Phys.\ Rev.\  {\bf 55}, 374 (1939),
  doi: 10.1103/PhysRev.55.374.

\bibitem{Tolman:1939jz} 
  R.~C.~Tolman,
  Phys.\ Rev.\  {\bf 55}, 364 (1939),
  doi:10.1103/PhysRev.55.364.


\bibitem{bps}
  G. Baym, C. Pethick, and P. Sutherland, 
  Astrophys. J. {\bf 170}, 299 (1971).


\bibitem{Grill:2014aea} 
  F.~Grill, H.~Pais, C.~Providência, I.~Vidaña and S.~S.~Avancini,
  Phys.\ Rev.\ C {\bf 90}, no. 4, 045803 (2014),
  doi:10.1103/PhysRevC.90.045803,
  [arXiv:1404.2753 [nucl-th]].


\bibitem{Ozel:2015fia} 
  F.~Ozel, D.~Psaltis, T.~Guver, G.~Baym, C.~Heinke, and S.~Guillot,
  Astrophys.\ J.\  {\bf 820}, no. 1, 28 (2016),
  doi:10.3847/0004-637X/820/1/28,
  [arXiv:1505.05155 [astro-ph.HE]].



\bibitem{Steiner:2015aea} 
  A.~W.~Steiner, J.~M.~Lattimer, and E.~F.~Brown,
  Eur.\ Phys.\ J.\ A {\bf 52}, no. 2, 18 (2016),
  doi:10.1140/epja/i2016-16018-1,
  [arXiv:1510.07515 [astro-ph.HE]].

\bibitem{suleimanov16}
V.F. Suleimanov, J. Poutanen, D. Klochkov, and K. Werner, 
Eur. Phys. J. A {\bf 52}, 20 (2016)

\bibitem{Chen:2015zpa} 
  W.~C.~Chen and J.~Piekarewicz,
  Phys.\ Rev.\ Lett.\  {\bf 115}, no. 16, 161101 (2015),
  doi:10.1103/PhysRevLett.115.161101,
  [arXiv:1505.07436 [nucl-th]].

\bibitem{haensel16}
P. Haensel, M.  Bejger, M. Fortin, and L. Zudnik, 
Eur. Phys. J. A {\bf 52}, 59 (2016).


\end{thebibliography}

\end{document}